\documentclass[twocolumn]{aastex631}
\usepackage{multirow}
\usepackage{makecell}
\usepackage{lipsum} 
\usepackage{xcolor} 
\usepackage{appendix}

\newcommand{\red}[1]{{\color{red}#1}}

\newcommand{\blue}[1]{{\color{blue}#1}}

\shorttitle{The energy-dependent temporal nature of MAXI J1803-298}
\shortauthors{Pradhan et al.}

\begin{document}

\title{ Probing the energy-dependent temporal nature of MAXI J1803-298 with \emph{AstroSat} and \emph{NICER}}  

\author{Arbind Pradhan}
\affiliation{Department of Applied Sciences, Tezpur University, Napaam, Assam 784028, India}
\author{Akash Garg}
\affiliation{Inter-University Centre for Astronomy and Astrophysics (IUCAA), Pune, Maharashtra 411007, India}
\author{Ranjeev Misra}
\affiliation{Inter-University Centre for Astronomy and Astrophysics (IUCAA), Pune, Maharashtra 411007, India}
\author{Biplob Sarkar}
\altaffiliation{E-mail: biplobs@tezu.ernet.in}
\affiliation{Department of Applied Sciences, Tezpur University, Napaam, Assam 784028, India}






\begin{abstract}

We performed the spectral and temporal analysis of MAXI J1803-298 using \emph{AstroSat}/LAXPC and \emph{NICER} observations taken in May 2021 during the initial phase of the outburst. We found that the source traverses through the hard, intermediate, and soft spectral states during the outburst. The spectrum in all states can be described using soft emissions from the thermal disk and hard emissions from the coronal regions. The variation in the inner disk temperature and normalization of the disk indicates the motion of the truncated disk across these different spectral states. We confirmed the presence of broad features, Type-C, and Type-B QPOs in the power spectra of different spectral states. We investigated the fractional rms and lags of all the variability features and discovered that the lag swung between positive and negative during the outburst evolution. While modeling the features with a simple model that considers variations in accretion parameters such as the accretion rate, heating rate, and inner disk radius, along with delays between them, we found a dynamic reversal in the origin of variability between the corona and the disk. Furthermore, our results are consistent with previous works and a radio study conducted on this source during its outburst.
\end{abstract}


\section{INTRODUCTION}\label{sec:intro}

X-ray binaries (XBs) are binary systems comprising a compact object and a companion star that prominently emit in the X-ray part of the electromagnetic spectrum. The compact object can either be an accreting black hole (BH) or a neutron star (NS). Depending on the mass of the companion star, XBs are classified as low-mass X-ray binaries (LMXBs) or high-mass X-ray binaries (HMXBs) \citep{1997xrb..book.....L}. HMXBs are mostly persistent, while LMXBs are transient and often remain in a low flux, quiescence state. Once an LMXB undergoes an outburst, the flux will increase, remaining high for weeks to months before slowly declining. This increase in luminosity is due to the increase in the accretion rate onto the compact object.

Most known BH XBs (or BHXBs) are transient in nature. During the outburst, a BHXB evolves through different spectral states before going back to its quiescent state and traces a `Q' shaped trajectory in the hardness intensity diagram \citep[HID;][]{2005Ap&SS.300..107H,2005A&A...440..207B,2012A&A...542A..56N}. The trajectory follows a sequence: Low-Hard State (LHS)-Hard Intermediate State (HIMS)-Soft Intermediate State (SIMS)-High-Soft State (HSS), then reverses through SIMS-HIMS-LHS \citep{2016ASSL..440...61B}. This transition exhibits hysteresis where high flux causes a transition from LHS to HSS, and then as the flux gets low, the source transits back into the LHS \citep{1995ApJ...442L..13M,2015A&A...574A.133K}.

Typically, the outburst starts with a faint brightness, and the X-ray emission is dominated by a hard power law of $\Gamma \sim$ 1.5 in LHS, accompanied by steady radio jets due to synchrotron emission. As the X-ray emission starts getting dominated by the thermal emission from the disk, the source proceeds towards HIMS and SIMS. Once the X-ray emission is prominently dominated by the soft photons, the HSS will be achieved. In the midst of this transition from LHS to HSS, radio emission quenches but recovers back during the transition to LHS \citep{2010MNRAS.403...61D}. For the transition from LHS to HSS, there are discrete ejections for collimated jets. According to the truncated disk model, the accretion disk is initially truncated in LHS, with soft photons from the disk being inversely Comptonized in the hot region near the BH, leading to a non-thermal power-law dominated photon spectrum \citep{1980A&A....86..121S,1994ApJ...434..570T}. Further, as the source moves to the soft state, the truncated disk is believed to move towards the innermost stable circular orbit (ISCO) and emits the thermal spectrum prominently.  Such a thermal emission is often described using the multicolor disk blackbody model \citep{1984PASJ...36..741M}. However, in the intermediate states, the spectrum consists of both thermal and non-thermal components. Additionally, there is another component that arises from the interaction between the hard X-rays and the disk, known as the reflection component, which can be seen as the ﬂuorescent and reprocessed emission such as the broad, asymmetric Fe K$\alpha$ line between 6 and 7 keV \citep{1991ApJ...376...90L} and the Compton hump above $\sim$ 15 keV \citep{2007ARA&A..45..441M}.

XBs exhibit variability in their light curves in the form of broadband noise (BBN) and aperiodic oscillations known as quasi-periodic oscillations (QPO) \citep{2005AN....326..798V}. Using Fourier techniques, BBN and QPO can be detected as a broad and narrow peak in the power density spectrum (PDS). It is crucial to understand the QPOs, as they are believed to be produced from the proximity of a compact object and can thereby reveal the behavior of the matter in the vicinity of a strong gravitational field. QPOs are broadly classified as low-frequency QPOs (a few mHz to $\sim$ 30 Hz) and high-frequency QPOs \citep[also known as kHz QPOs in neutron star XBs (NSXBs);][]{2016AN....337..398M}. High frequency QPOs have also been reported in BHXBs such as GRS 1915+105 \citep{2013MNRAS.432...10B} and IGR J17091-3624 \citep{2012ApJ...747L...4A}. On the basis of intrinsic properties like centroid frequency, width, and strength, low-frequency QPOs are further classified as types A, B, and C. Interestingly, different spectral states exhibit different types of low-frequency QPO \citep{wijnands1999complex,homan2001correlated,casella2004study}, indicating a correlation between the spectral and temporal behavior of XBs.

Although several models have been proposed to understand the physics of QPOs (mainly type-C QPOs), there is still no consensus on the exact origin of the QPOs. These models are either based on the geometrical effect or thermal instabilities in the accretion disk. For instance, models like the Relativistic Precession Model (RPM) \citep{stella1999correlations, ingram2009low} are based on the geometric origin, while models like the transition layer model \citep{titarchuk1999correlations,titarchuk2004spectral} are based on instability leading to the production of low frequency QPOs. Along with these models there are models based on the shock oscillation causing the low frequency QPOs in BHXBs such as \cite{1999ApJMol,2000ApJCha}. There is an alternate approach based on the observed correlations between spectral and temporal properties of XBs. It has been confirmed in many sources that the QPO centroid frequency has strong correlations with spectral parameters (e.g. \cite{garg2022energy,2023MNRAS.524.2721D} and references therein), indicating a connection between the radiative properties of accretion and QPOs. The QPOs possess energy-dependent properties such as fractional root mean square (rms) and time lag, which can be directly connected with spectral variations. These properties can help identify the radiative mechanisms responsible for the presence of QPOs. \citet{lee2001compton}, \citet{kumar2014energy}, and \citet{karpouzas2020comptonizing} proposed models based on the idea of the Comptonized photons impinging back onto the seed photon source to describe the occurrence of hard and soft time lags for kHz QPO in NSXBs. Furthermore, based on the same framework, \citet{karpouzas2021variable}, \citet{bellavita2022vkompth}, and \citet{garcia2022evolving} explained the observed energy-dependent rms and lag for low-frequency QPOs in BHXB. \cite{misra2013alternating}, \cite{mir2016model} and \citet{maqbool2019stochastic} considered variations in spectral parameters like accretion rate, inner disk radii, and power law index, along with propagation delays between them, to model the rms and time delays for the QPOs in NS and BH systems.

\cite{garg2020identifying} formulated a generic technique to model the timing behavior of variability using the spectral information of the accretion flow. The technique requires modeling the time-averaged photon spectra to determine spectral components such as the disk blackbody and thermal Comptonization of the X-ray emission. Further, in the approach, the spectral parameters are translated into their physical counterparts to determine which variations in physical parameters can reproduce the observed rms and lag associated with QPO or BBN. \cite{garg2020identifying} applied their model to study the temporal nature of GRS 1915+105. Similarly, \cite{garg2022energy} studied the QPOs observed in MAXI J1535-371 using the model and found that the variations in the mass accretion rate, inner disk radius, and heating rate, along with delays between them could properly fit the rms and time lag spectra. \cite{husain2023investigating} also used this technique to investigate the QPOs and harmonics observed in H 1743-322. Recently, \cite{2024Hitesh} and \cite{2024Dhaka} also applied the technique to the energy-dependent rms and time lags of broad features, often observed in the PDS of BH systems, GX 339–4 and GRS 1915+105, respectively. This technique has been successfully applied to NSXBs, in sources such as 4U 1608–52 \citep{2024Sree} and GX 9+9 \citep{2024GX9+9}.

\begin{figure*}[!ht]
    \centering
    \centering
    \includegraphics[width=0.5\textwidth,keepaspectratio=true]{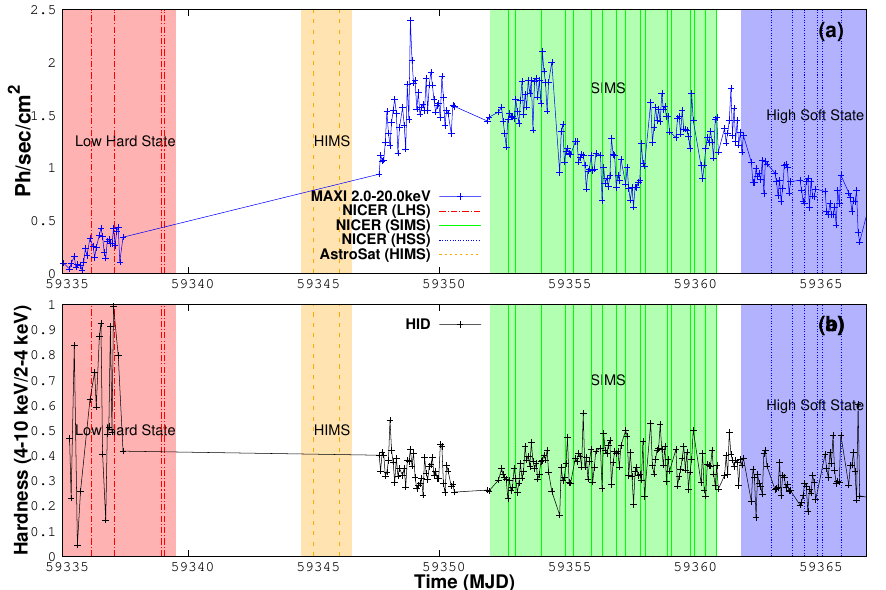}
\includegraphics[width=0.49\textwidth, height=6.5cm]{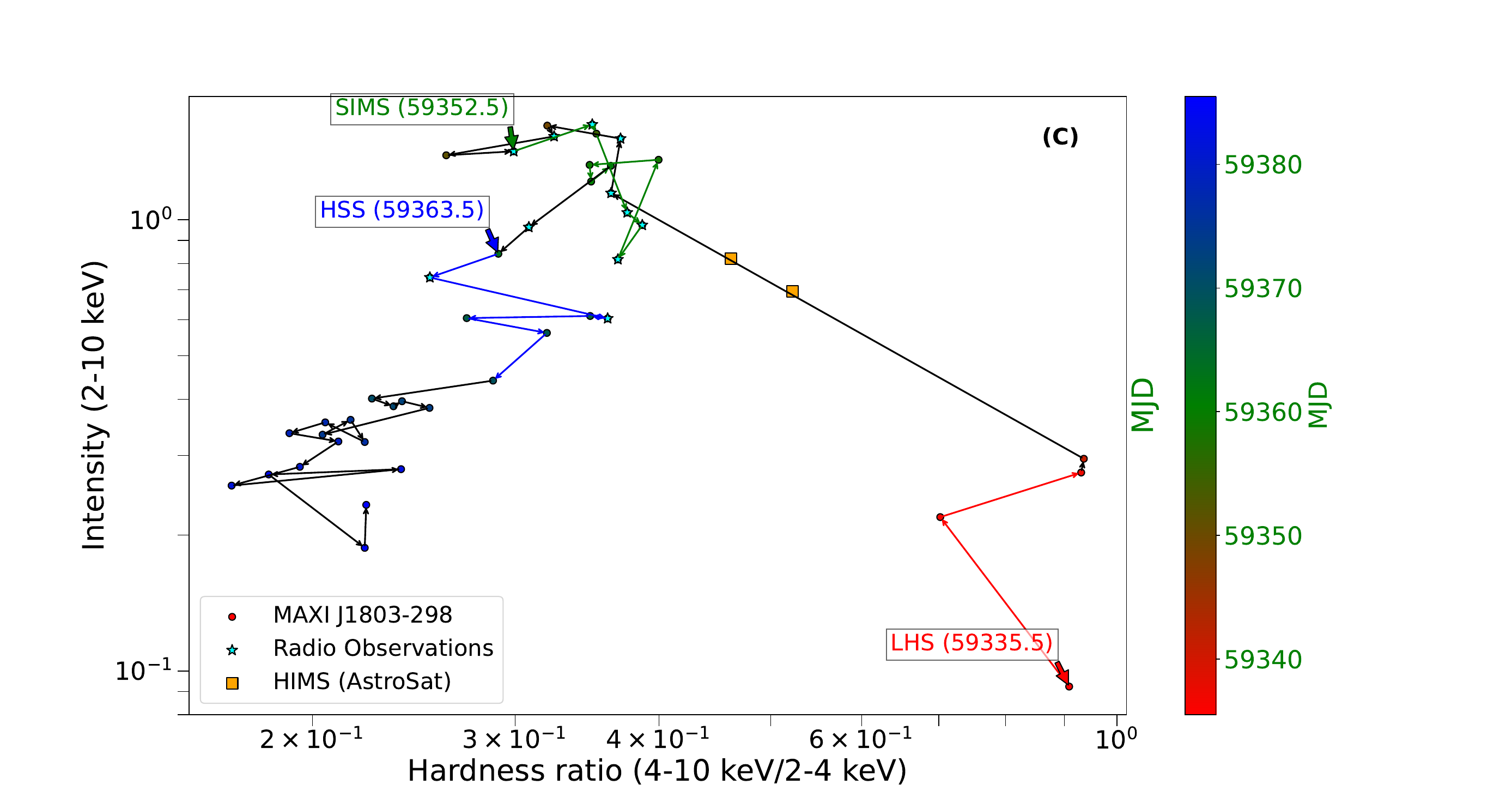}
\caption{The figure (a) shows the MAXI light curve of MAXIJ1803 in 2-20 keV. The vertical lines represent observations taken by \emph{NICER} (red, green, and blue) and \emph{AstroSat} (orange). The different colors represent different spectral states: LHS (red), HIMS (orange), SIMS (green), and HSS (blue). The figure (b) shows hardness variations vs. time with hardness defined as 4-10 keV/2-4 keV. While figure (c) is the Hardness-Intensity Diagram (HID) of MAXIJ1803. The horizontal axis represents the hardness ratio (4-10 keV/2-4 keV), and the vertical axis is the intensity in 2-10 keV. The different colors represent different spectral states, and the stars on the plot represent the radio emission as taken from \citet{2023MNRAS.522...70W}. The different MJD box markers show the start of the indicated spectral state.}
    \label{fig:Sub-figure 1}
    \end{figure*}

The BH candidate MAXI J1803-298 (hereafter, MAXIJ1803) is one of the recently discovered X-ray transients during its outburst on May 1, 2021, by \textit{Monitor of All-sky X-ray Image/Gas Slit Camera} (\textit{MAXI/GSC}) \citep{2021ATel14587....1S}. The optical counterpart of the source was detected by \emph{Swift} \citep{2021ATel14591....1G, 2021ATel14594....1H}. Optical \textit{Southern African Large Telescope} (\textit{SALT}) spectroscopy suggested the presence of the accreting BH in MAXIJ1803 \citep{2021ATel14597....1B}. The multi-wavelength follow-up of MAXIJ1803 was performed by \cite{Mata_2022}, which also suggested the compact object to be a BH with a mass of 3-10 $M_\odot$. During its initial outburst, the detection of periodic absorption dips indicated the source with high inclination ($\gtrsim$70$^{\circ}$) and the periodicity was of $\sim$ 7 hours \citep{2021ATel14606....1H, 2021ATel14609....1X, Jana_2022}. Radio wavelength was detected in the initial outburst phase with \emph{MeerKAT} \citep{2021ATel14607....1E}. \cite{2022ApJ...933...69C} stated that the source was in HIMS with detection of type-C QPO and estimated BH mass of $\sim$ 8-16 $M_\odot$ with a spin `$a$' of $\gtrsim$ 0.7 using the \emph{Astrosat} observation. Later on, \cite{Jana_2022} also supported the transitioning from HIMS to SIMS using the \emph{Astrosat} observations. \textit{MAXI/GSC} monitored the source during its whole outburst, but it was unable to observe MAXIJ1803 for eight days from May 4 (MJD 59338) to May 12 (MJD 59346). \citet{Shidatsu2022} studied the X-ray spectral evolution with \textit{MAXI/GSC} and \textit{Swift/BAT} data and showed the transition from LHS to HSS. MAXIJ1803 remained in the outburst phase till October 19 (MJD 50906) \citep{2021ATel14994....1S}. Based on \emph{Nuclear Spectroscopic Telescope Array (NuSTAR)} and \emph{the Neutron Star Interior Composition Explorer (NICER)} data, \cite{2022MNRAS.516.2074F} estimated the spin to be $\sim$ 0.991 with an inclination $`i'$ of $\sim$70$^{\circ}$. \cite{2023ApJ...949...70C} also constrained the inclination of the accretion disk to $\sim$ 75$^{\circ}$ using \emph{NuSTAR} data and reported type-B and type-C QPOs. \cite{2023MNRAS.523.4394Z} reported type-C QPOs using \emph{Insight}-HXMT data and type-B QPOs using \emph{NICER} data in different spectral states. Recently, \cite{2024Adegoke} studied the source in its LHS, intermediate, and HSS using \emph{NuSTAR} and \emph{NICER} and reported the presence of disk wind and dips in light curve due to the presence of photo-electric absorption.

Here, we intend to perform spectral and temporal analysis of the source in all its transitioned spectral states using the \emph{NICER} and \emph{AstroSat}/LAXPC observations taken during the month of May 2021. \emph{NICER} observations are in LHS (represented by HS segment), SIMS (represented by SIMS1, SIMS2, SIMS3, and SIMS4 segments), and HSS (represented by HSS segment), and to fill the gap of HIMS, we are using \emph{AstroSat} observations (denoted by HIMS segment). 
We are considering these two missions because of their excellent timing resolutions. The results of spectral analysis are used to model the energy-dependent properties with \cite{garg2020identifying,garg2022energy} to understand the QPOs and the aperiodic humps observed in the PDS. Furthermore, we will try to understand the cause and the propagation of the variability based on the results of modeling by \cite{garg2020identifying, garg2022energy}. Figure \ref{fig:Sub-figure 1} illustrates the profile of MAXIJ1803 during its outburst using MAXI observations. The source can be seen to transit through different spectral states, which we represent using different colors. In the following sections, we describe how we identify these different spectral states using spectral analysis, types of QPO observed in these states, and information from the previous works \citep[see][]{Mata_2022, Shidatsu2022, 2023MNRAS.522...70W}. Throughout this paper, we follow the color codes used in Figure \ref{fig:Sub-figure 1} to denote different spectral states. 

The structure of this paper is as follows: In Section \ref{sec:ODP}, we discuss the observation and data reduction. Section \ref{sec:DA} presents the detailed spectro-timing analysis and the modeling of the energy-dependent rms and time lag spectra. We have interpreted the results obtained from our analysis in Section \ref{sec:DC}.

\section{OBSERVATION AND DATA REDUCTION} \label{sec:ODP}
\subsection{AstroSat  Observation}
\emph{AstroSat} is India's first multi-wavelength astronomical satellite, which performs broadband coverage from optical to hard X-rays \citep{2006AdSpR..38.2989A,2022hxga.book...83S}. It consists of five payloads, namely Ultra-violet Imaging Telescopes (UVIT), Large Area X-ray Proportional Counters (LAXPC), Soft X-ray Telescope (SXT), Cadmium–Zinc–Telluride Imager (CZTI), and Scanning Sky Monitor (SSM).  Among its five payloads, we have used data from Large-area X-ray Proportional Counters \citep[LAXPC,][]{2017ApJS..231...10A, 2016SPIE.9905E..1DY} and Soft X-ray Telescope \citep[SXT,][]{2017JApA...38...29S, singh2021observations, bhattacharyya2021science}.

Both LAXPC and SXT took two Target of Opportunity (ToO) observations of MAXIJ1803 from May 11 at 01:09 UTC to May 11 at 11:59 UTC (hereafter, we call this A1) and from May 11 at 12:05 UTC to May 12 at 13:03 UTC (hereafter, we call this A2). Table \ref{tab:table1} gives the observation IDs and exposure time details.

\subsubsection{LAXPC Data Reduction}

Among the three LAXPC units, we used only LAXPC20 data in our analysis, as LAXPC10 has a low gain issue, while LAXPC30 is not functioning currently. We used the standard analysis software \texttt{LAXPCsoftware}\footnote{\url{http://astrosat-ssc.iucaa.in/laxpcData}} (\texttt{LAXPCSOFT}; version 2022 August 15) to process level 1 data into level 2 data. We merged both the observations (A1 and A2) using \texttt{"ftmerge"} and generated a single clean Eventfile. Further, we generated light curves and spectra using the standard sub-routines available in \texttt{LAXPCsoftware}\footnote{\url{http://astrosat-ssc.iucaa.in/laxpcData}}. 



\subsection{NICER  Observation}\label{sec:NICER_OBS}
The \emph{NICER} mission was launched in 2017; since then, it has been mounted on the International Space Station (ISS). It has good spectral and timing capabilities with a large collecting area at $\sim$1 keV and a timing resolution of 300 nanoseconds \citep{gendreau2017searching}. X-ray timing instrument (XTI; \citep{gendreau2016neutron}) onboard \emph{NICER} consists of 56 co-aligned X-ray concentrator optics and silicon drift detector pairs. It provides an effective area of 1900 cm$^2$ at $\sim$1.5 keV and an energy resolution of $\sim$ 100 eV. The sensitive energy range of \emph{NICER} is 0.2 - 12 keV \citep{arzoumanian2014neutron}.

\emph{NICER} observed MAXIJ1803 almost through the full outburst from May 2 till November 7, 2021. In this work, we have taken 27 \emph{NICER}  observations made during May 2021, with no observations between May 6 and May 17. 
We analyzed these \emph{NICER} datasets using \emph{NICER} data analysis software (NICERDAS) (version 2023-08-22\_V011a) and CALDB (version 20240206\footnote{\url{https://heasarc.gsfc.nasa.gov/docs/heasarc/caldb/nicer/}}). We reduced the \emph{NICER} data using standard \texttt{nicerl2} pipeline\footnote{\url{https://heasarc.gsfc.nasa.gov/lheasoft/ftools/headas/nicerl2.html}}. 
Using \texttt{nicerl3-spect\footnote{\url{https://heasarc.gsfc.nasa.gov/docs/software/lheasoft/help/nicerl3-spect.html}}} task we extracted the spectra, rmf and arf. \texttt{nibackgen3C50} was chosen for the spectral background. The spectra generated by the \texttt{nicerl3-spect} task, by default, include a systematic uncertainty, which was retained during the spectral analysis. Additionally, we used \texttt{NICER\_RM\_Software}\footnote{The \texttt{NICER\_RM\_Software} is a package developed by Prof. R. Misra (IUCAA) and consists of subroutines similar to LAXPC software, to analyze \texttt{NICER} data.} for extracting light curves, power spectra, rms, and time lag spectra.

Figure \ref{fig:Sub-figure 1} shows the MAXI long-term light curve for May 2021 generated using the orbit data extracted from \texttt{maxi.rixen} website\footnote{\url{http://maxi.riken.jp/star_data/J1803-298/J1803-298.html}}. The dashed and solid vertical lines represent the \emph{AstroSat} and \emph{NICER} observations of the source, respectively. The lower panel of Figure \ref{fig:Sub-figure 1} shows the evolution of the hardness of the source with time. The color codes represent different spectral states of the source during the outburst. \emph{AstroSat }  observed the source during HIMS \citep{2022ApJ...933...69C, Jana_2022}, whereas \emph{NICER}  covered LHS, SIMS, and HSS portions \citep{2023MNRAS.523.4394Z, zhang2024variable} of the outburst. Figure \ref{fig:Sub-figure 1} (c) shows the HID generated using MAXI one-day binned data with different spectral states marked using the same color codes as in Figure \ref{fig:Sub-figure 1}.  There were no data from 59341.5 to 59347.5 as MAXI was off during that period, so we have marked the \emph{AstroSat}( HIMS) observations in Figure \ref{fig:Sub-figure 1} (c) using the MJD of \emph{AstroSat} observations. The star markers in Figure \ref{fig:Sub-figure 1} (c) represent radio monitoring of MAXIJ1803 through intermediate and soft states, as carried out by \citet{2023MNRAS.522...70W} using the Very Long Baseline Array (VLBA) observations between 2021 May 13 (59347.41) and 2021 June 7 (59372.34) \citep[See Table 1 in ][]{2023MNRAS.522...70W}. Along with VLBA, \citet{2023MNRAS.522...70W} have utilized observations of the Atacama Large Millimeter/Sub-Millimeter Array (ALMA) taken on 2021 May 11 (59345.19) and 2021 May 15 (59349.24), and the observations of the Australia Telescope Compact Array (ATCA) taken between 2021 May 11 (59345) and 2021 July 3 (59398)). On analyzing ATCA observations, they found that the spectral index ($\alpha: (S_\nu \propto \nu^\alpha))$ of the radio spectrum at the beginning of the radio flare (59345.60 ± 0.17) was -0.1 $\pm$0.1. Then, in the next observation (59347.88 ± 0.03), they noticed the peak radio flux density and the steepest radio spectrum with $\alpha =$ -0.8 $\pm$ 0.2. The steepest spectral index marks the presence of an extended, optical thin transient radio jet in LMXB.  Further, they observed a gradual flattening of the radio spectrum and suggested the gradual transition of states from intermediate to HSS \citep[see Figure 2 of][]{2023MNRAS.522...70W}. Here in Figure \ref{fig:Sub-figure 1} (c), we are just denoting different states based on our analysis and previous works. The color bar represents the evolution of the different states with MJD.

We further divided the HS, HIMS, HSS, and the four SIMS segments into sub-parts to test whether the spectral and timing properties varied during these segments. However, we didn't find any significant variation in the spectral and temporal parameters and hence retained the present form of segmentation of the observations.
\begin{table*}[!ht]
    \caption{Log of the observations used in the analysis along with their exposure time.}
    \label{tab:table1}
     \scalebox{0.9}{
    \begin{tabular}{|c|c|r|r|r|}
     \hline
       Segments & Instruments  & Observation Ids  & \makecell{Date of Obs. \\ Start MJD ------ Stop MJD} & Exposure time (s) \\
        \hline
       \textbf{HS} &\textbf{\emph{NICER}} &  \begin{tabular}{@{}c@{}} 4202130101  \\ 4202130102 \\ 4202130103 \\ 4202130104 \end{tabular}  & \begin{tabular}{@{}c@{}} 59336.149 ------ 59336.867 \\ 59337.053 ------ 59337.060 \\ 59338.927 ------ 59338.995 \\ 59339.056 ------ 59339.96 \end{tabular}  & \begin{tabular}{@{}c@{}} 516.00  \\ 535.00 \\ 566.00 \\ 	4039.00 \end{tabular} \\ \hline 
      \multirow{2}{*}{\rotatebox{0}{\textbf{HIMS} }}&\multirow{2}{*}{\rotatebox{0}{\textbf{\emph{AstroSat}}}} &  T04 003T01 9000004368 \textbf{(A1)} & 59345.048 ------ 59345.499  & 40000.00 \\  
       &&  T04 006T01 9000004370 \textbf{(A2)} & 59345.504 ------ 59346.725 &  50000.00\\ \hline
       \textbf{SIMS1}&\textbf{\emph{NICER}} &   \begin{tabular}{@{}c@{}} 4202130105  \\ 4202130106 \\ 4202130107 \end{tabular}  & \begin{tabular}{@{}c@{}}  59352.745 ------ 59352.951 \\ 59353.006 ------ 59353.925 \\ 59354.038 ------  59354.501 \end{tabular}  & \begin{tabular}{@{}c@{}} 2680.00  \\ 12211.00 \\ 4631.00 \end{tabular} \\ \hline
       &  &4202130108  & 59355.334 ------ 59355.868 & 3809.00  \\ 
      & & 4202130109 & 59356.904 ------ 59356.496 & 3020.00  \\ 
      \textbf{SIMS2}& \textbf{\emph{NICER}} &  4675020103 & 59357.012 ------ 59357.938 & 5372.00  \\ 
       & & 4202130110 & 59357.401 ------ 59357.227 & 7508.00  \\ 
      & & 4202130111 & 59357.982 ------ 59358.058 & 2456.00 \\ 
       & & 4675020104 & 59358.175 ------ 59358.391 & 2671.00 \\ \hline
       \textbf{SIMS3}&\textbf{\emph{NICER}} &  \begin{tabular}{@{}c@{}} 4202130112 \\ 4675020105 \end{tabular}  &  \begin{tabular}{@{}c@{}} 59359.078 ------ 59359.941 \\ 59359.224 ------ 59359.475 \end{tabular} & \begin{tabular}{@{}c@{}} 3683.00 \\ 2535.00 \end{tabular}  \\ \hline
       \textbf{SIMS4} & \textbf{\emph{NICER}} &  \begin{tabular}{@{}c@{}} 4675020106 \\ 4642010101 \\ 4642010102 \end{tabular} &   \begin{tabular}{@{}c@{}} 59360.120 ------ 59360.459 \\ 59360.574 ------ 59360.967 \\ 59361.026 ------ 59361.225 \end{tabular}  & \begin{tabular}{@{}c@{}} 1867.00 \\ 4318.00 \\ 768.00 \end{tabular}   \\ \hline
       \textbf{HSS} &\textbf{\emph{NICER}} & \begin{tabular}{@{}c@{}} 4675020107  \\ 4202130114 \\ 4675020108 \\ 4675020109 \\ 4202130115 \\ 4675020110 \end{tabular}  &  \begin{tabular}{@{}c@{}} 59363.211 ------ 59363.991 \\ 59364.051 ------ 59364.443 \\ 59364.501 ------ 59364.960 \\ 59365.018 ------ 59365.993 \\ 59365.211 ------ 59365.670 \\ 59365.997 ------ 59369.942 \end{tabular} & \begin{tabular}{@{}c@{}} 4926.00  \\ 1749.00 \\ 3574.00 \\ 5859.00 \\ 3912.00 \\ 4791.00 \end{tabular}      \\ \hline
       \end{tabular}
       }

\end{table*}

\begin{figure*}[!ht]
    \centering
    \includegraphics[width=0.49\linewidth]{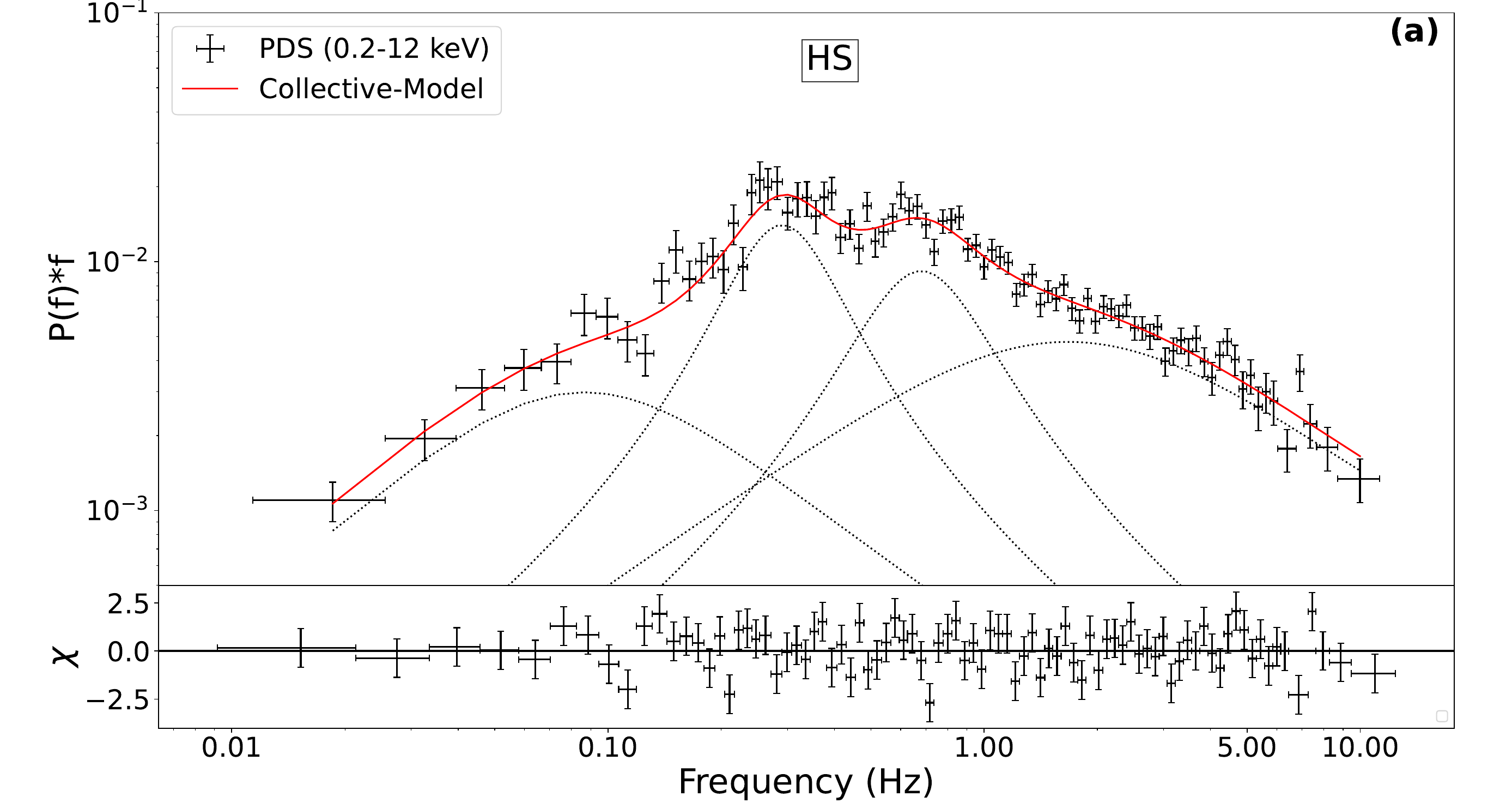}
        \includegraphics[width=0.49\linewidth]{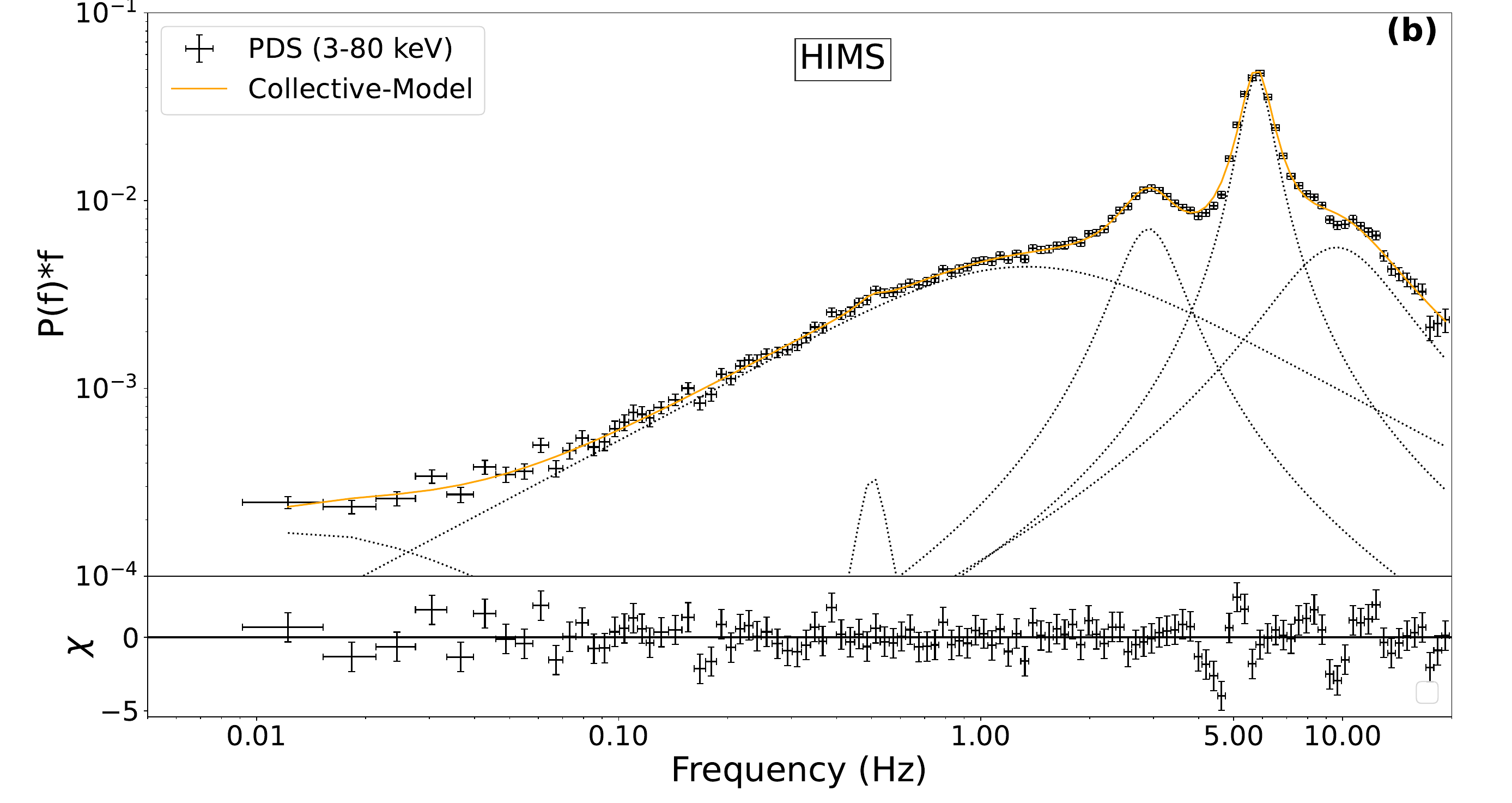}
    \includegraphics[width=0.49\linewidth]{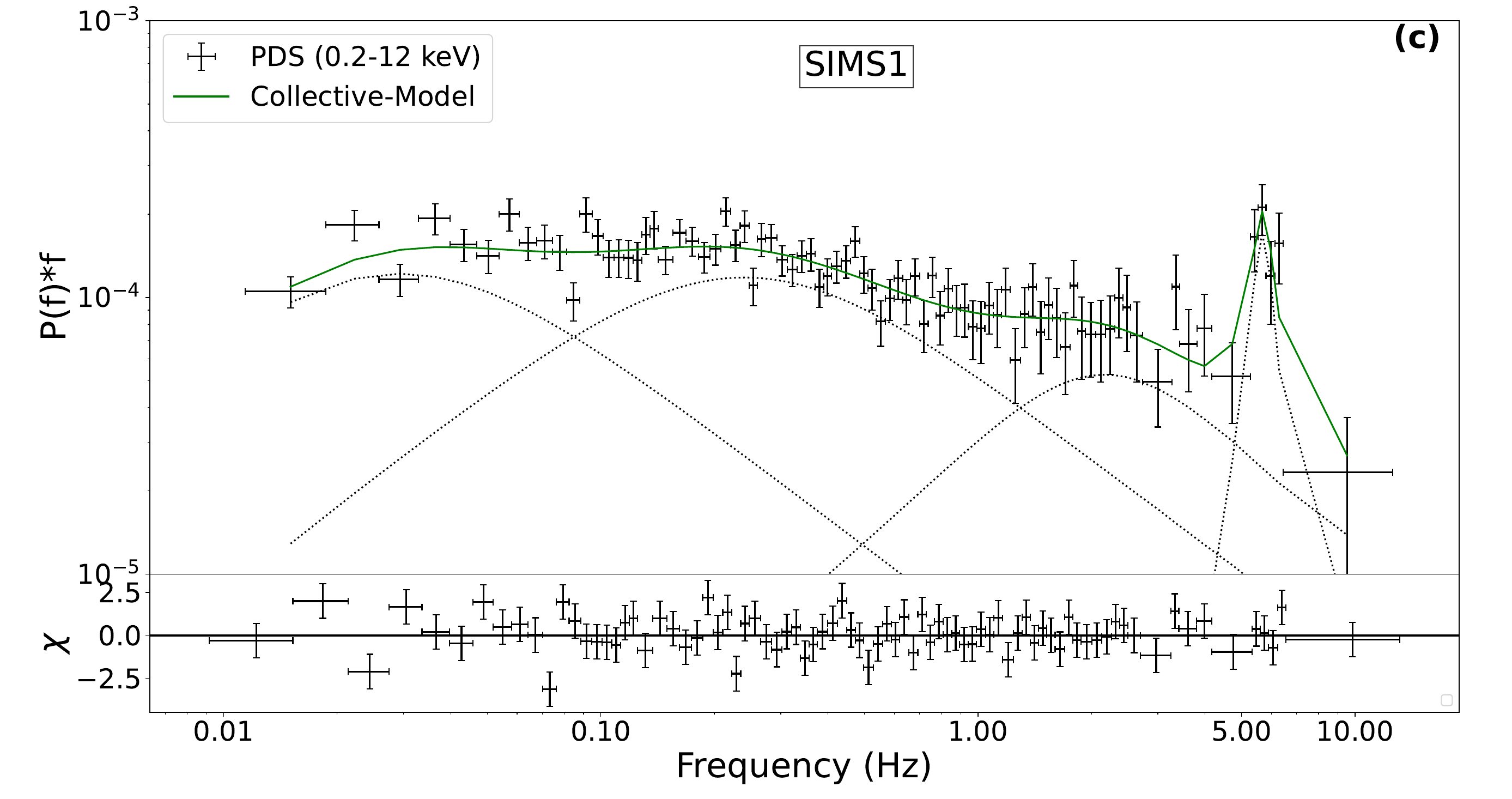}
    \includegraphics[width=0.49\linewidth]{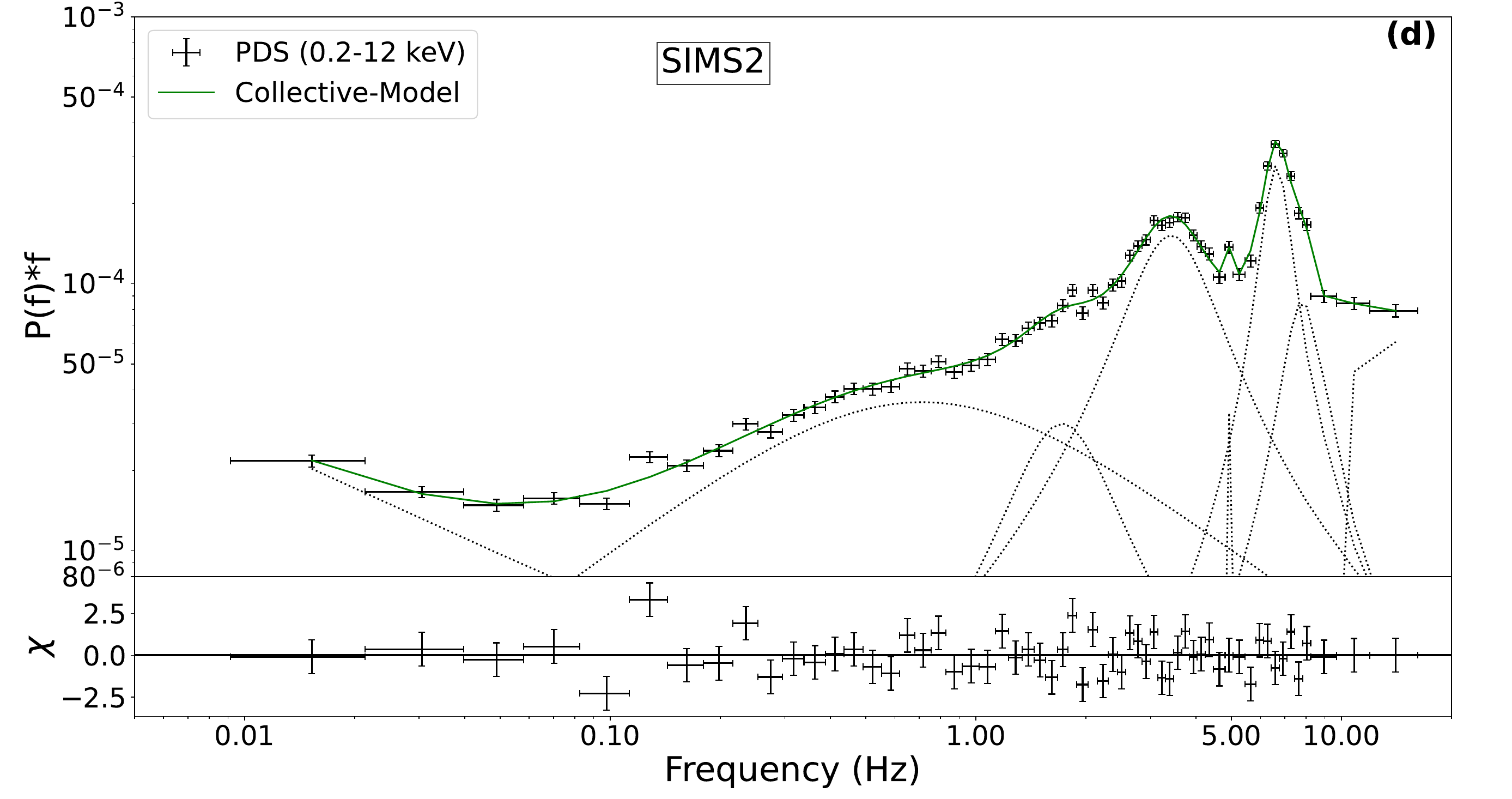}
    \includegraphics[width=0.49\linewidth]{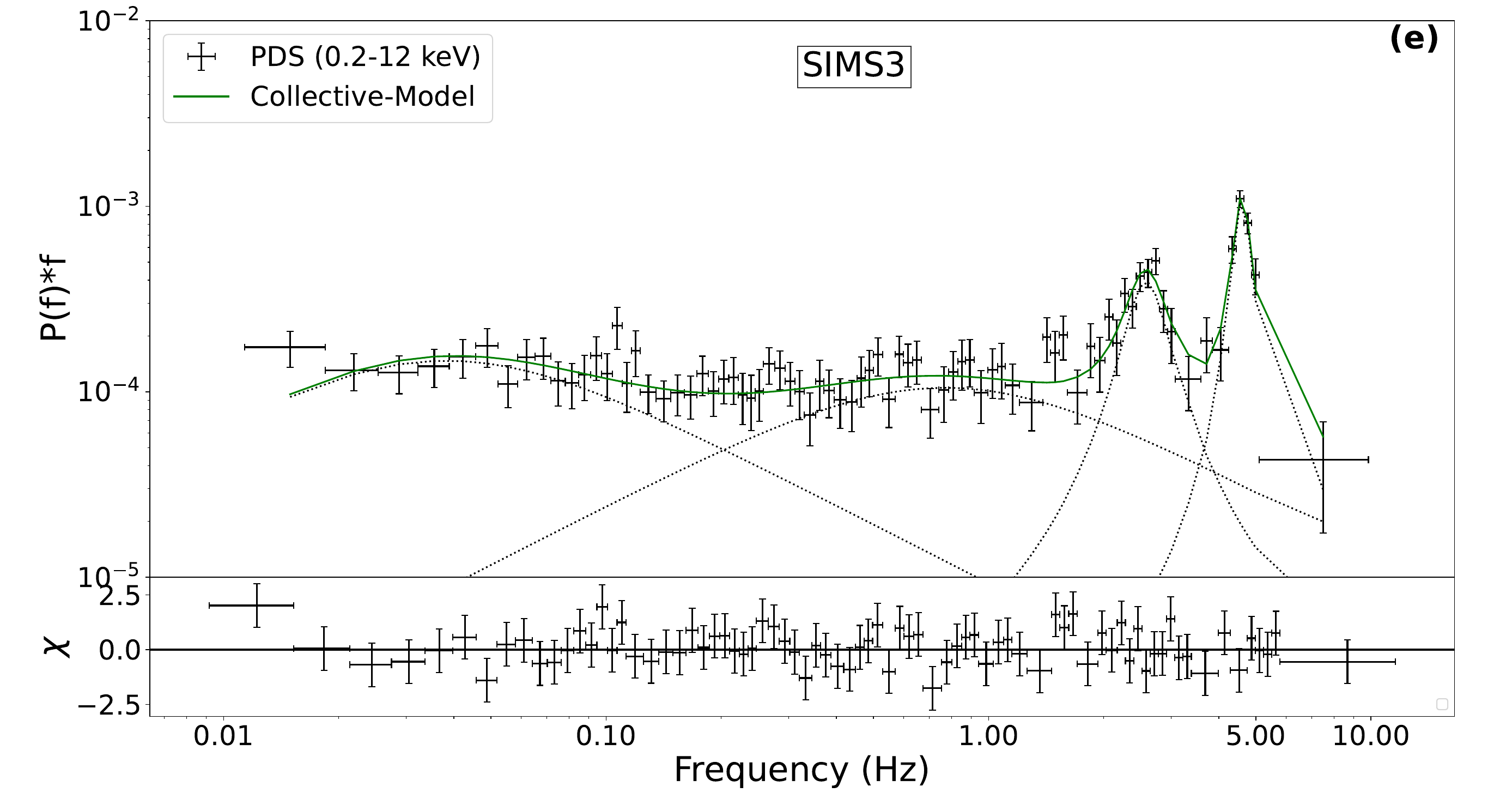}
    \includegraphics[width=0.49\linewidth]{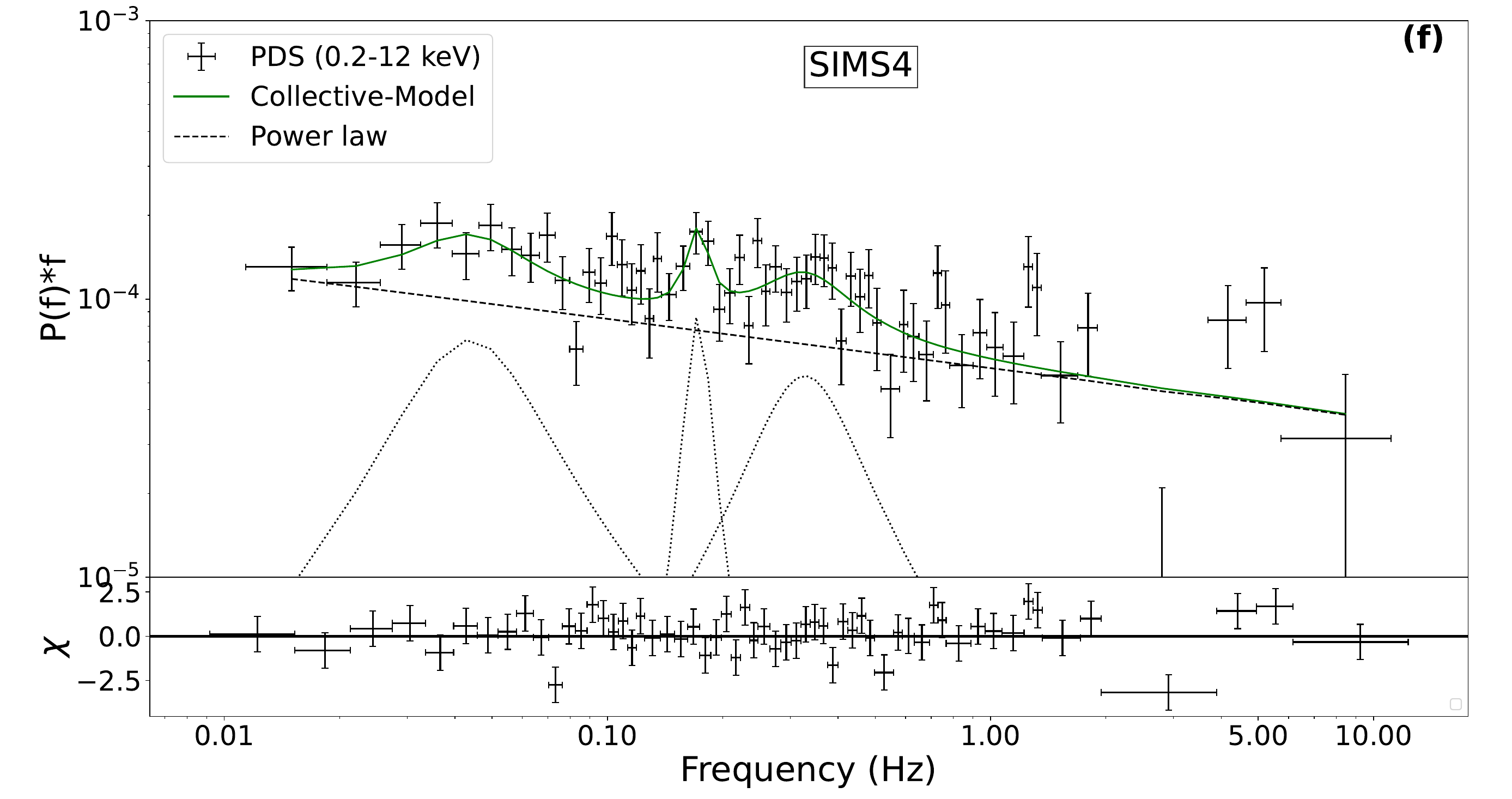}
    \includegraphics[width=0.5\linewidth]{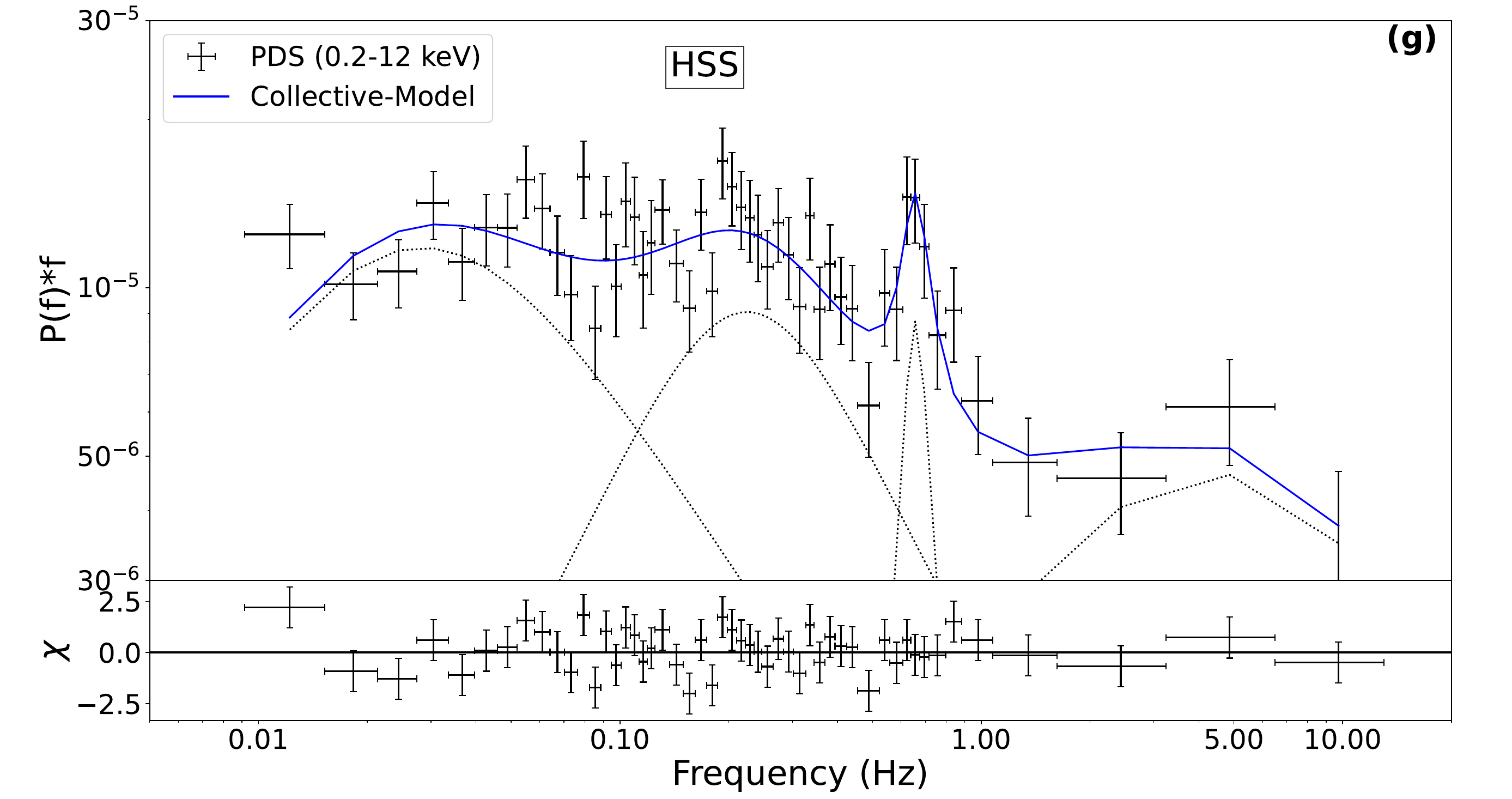}
\caption{Power Density Spectra (PDS) of all segments in the 0.005–20 Hz frequency range. Each panel shows the PDS of an individual segment fitted with multiple Lorentzian components, along with the corresponding residuals from the best-fit model.}    
\label{fig:PDS_seg_1_7}
\end{figure*}

\section{DATA ANALYSIS} \label{sec:DA}

    \begin{figure*}[!ht]
    \centering
    \includegraphics[width=5cm,angle=-90]{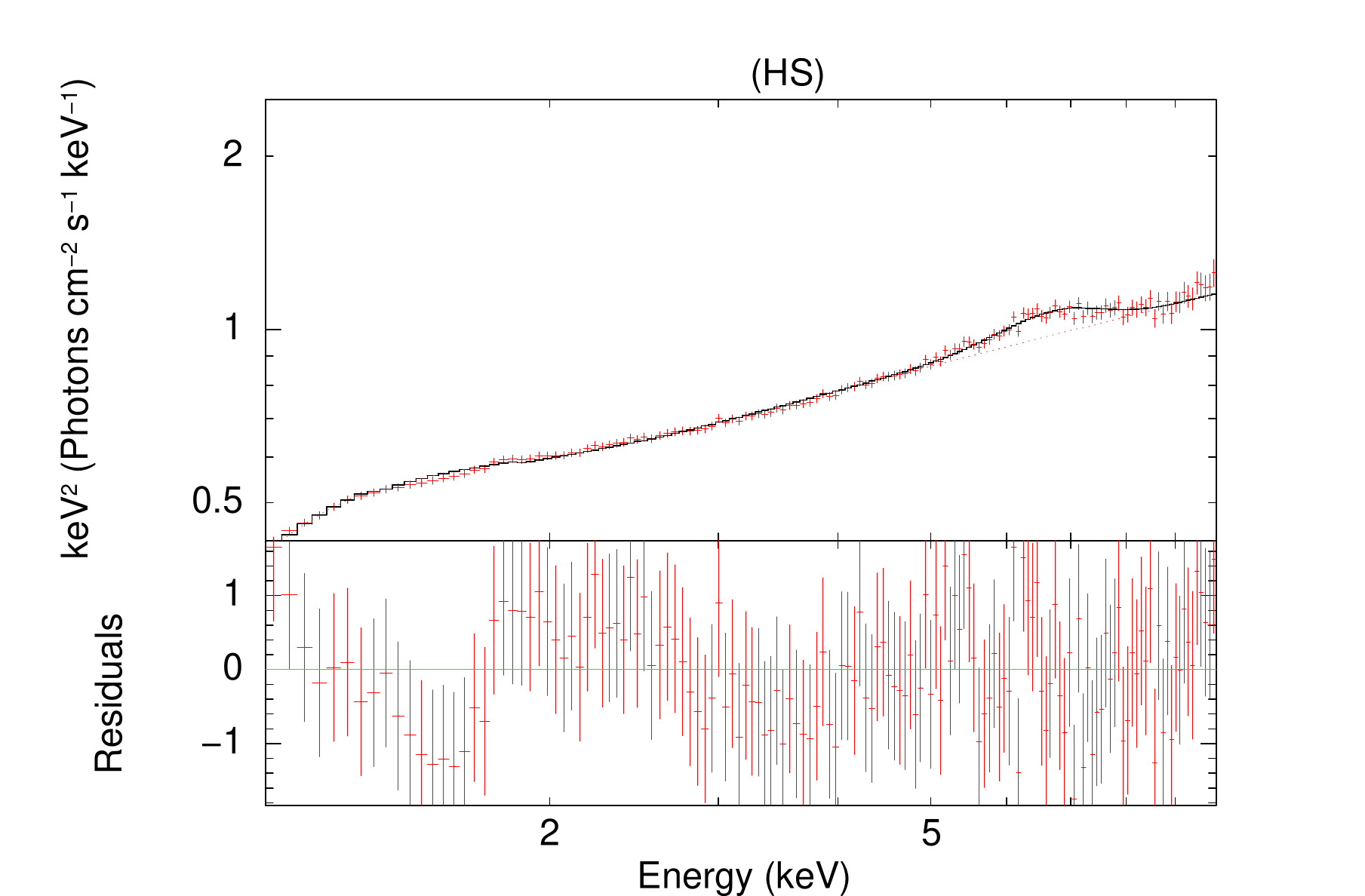}
    \includegraphics[width=5cm,angle=-90]{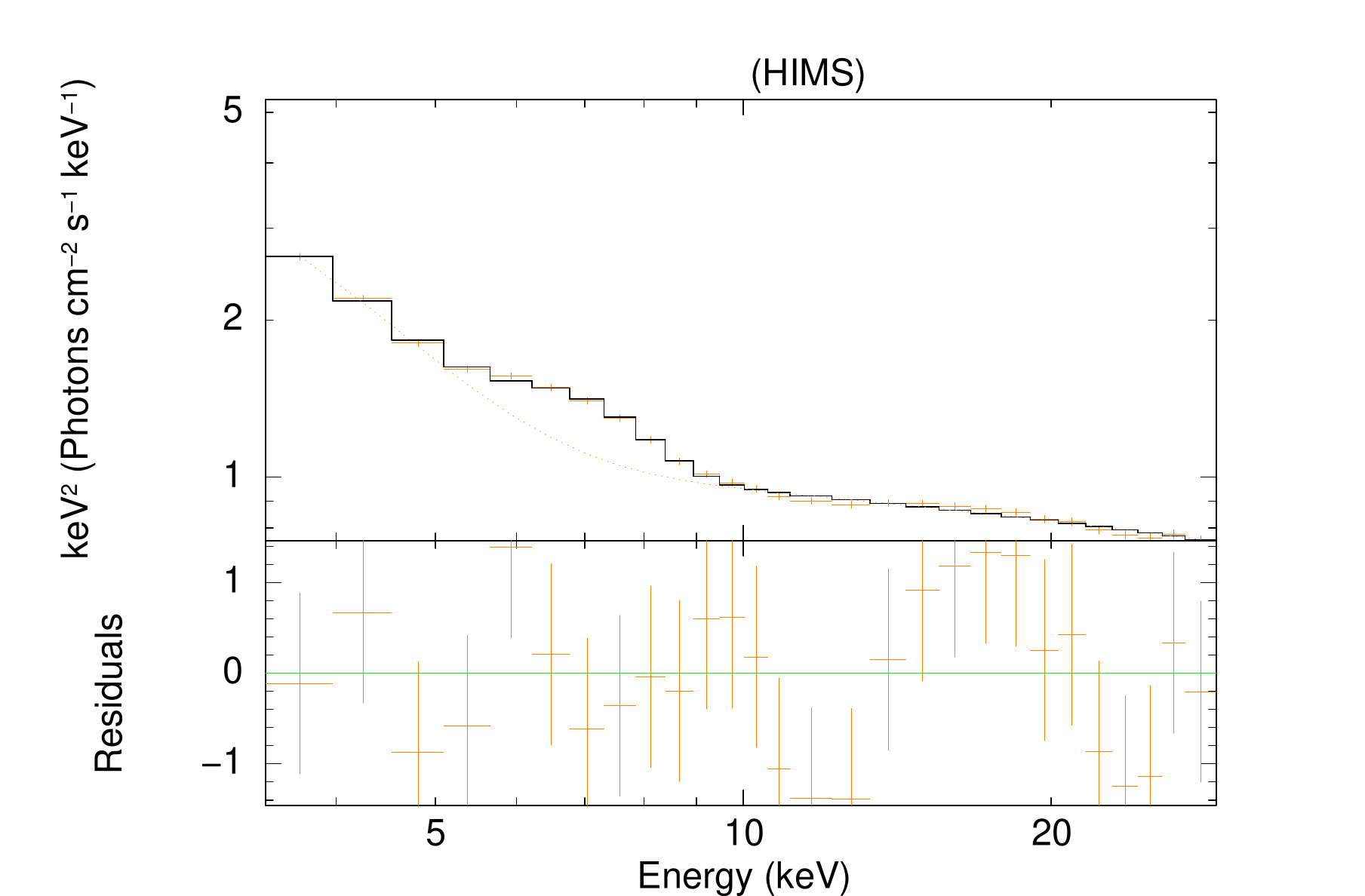}
    \includegraphics[width=5cm,angle=-90]{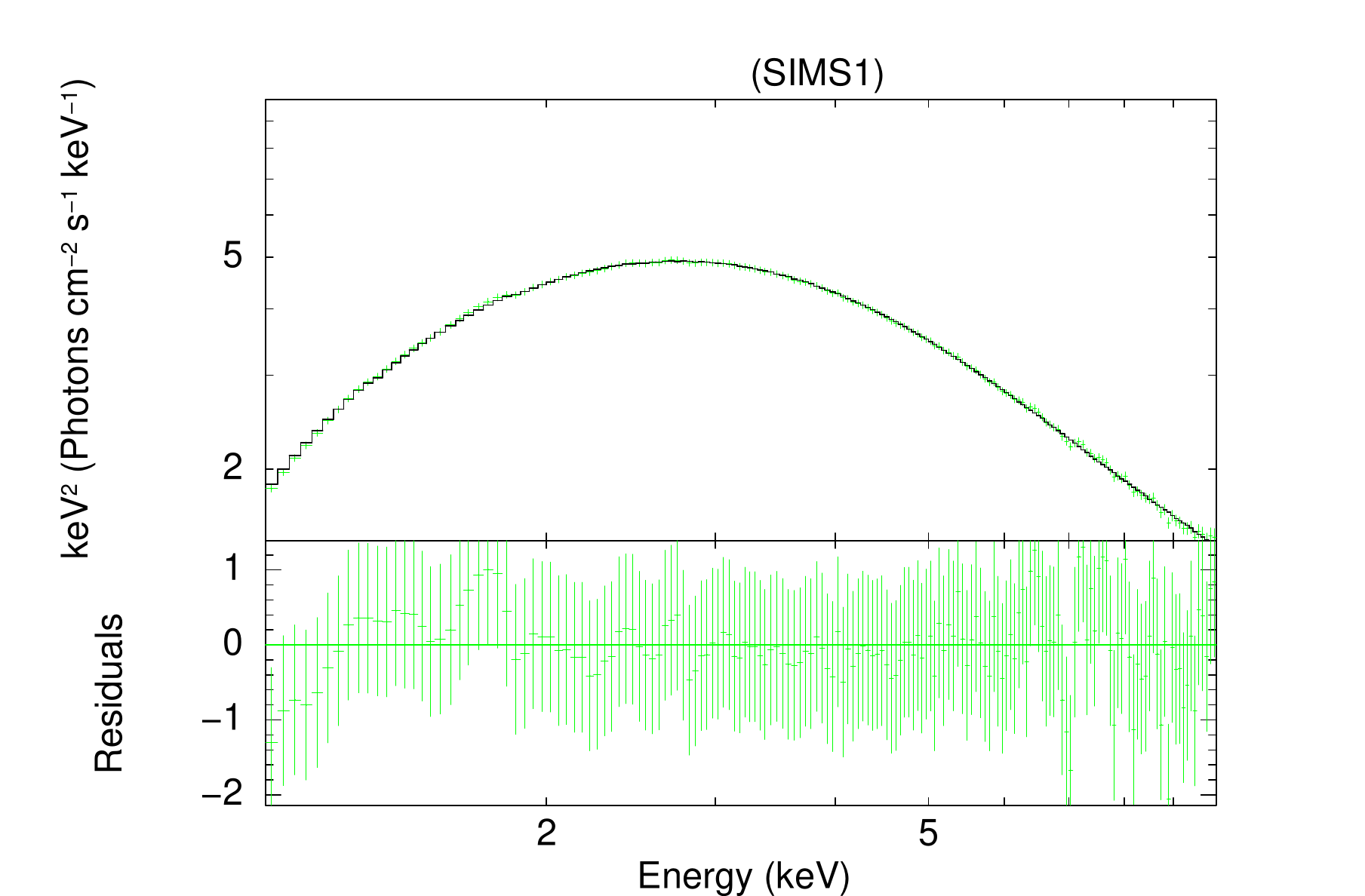}
    \includegraphics[width=5cm,angle=-90]{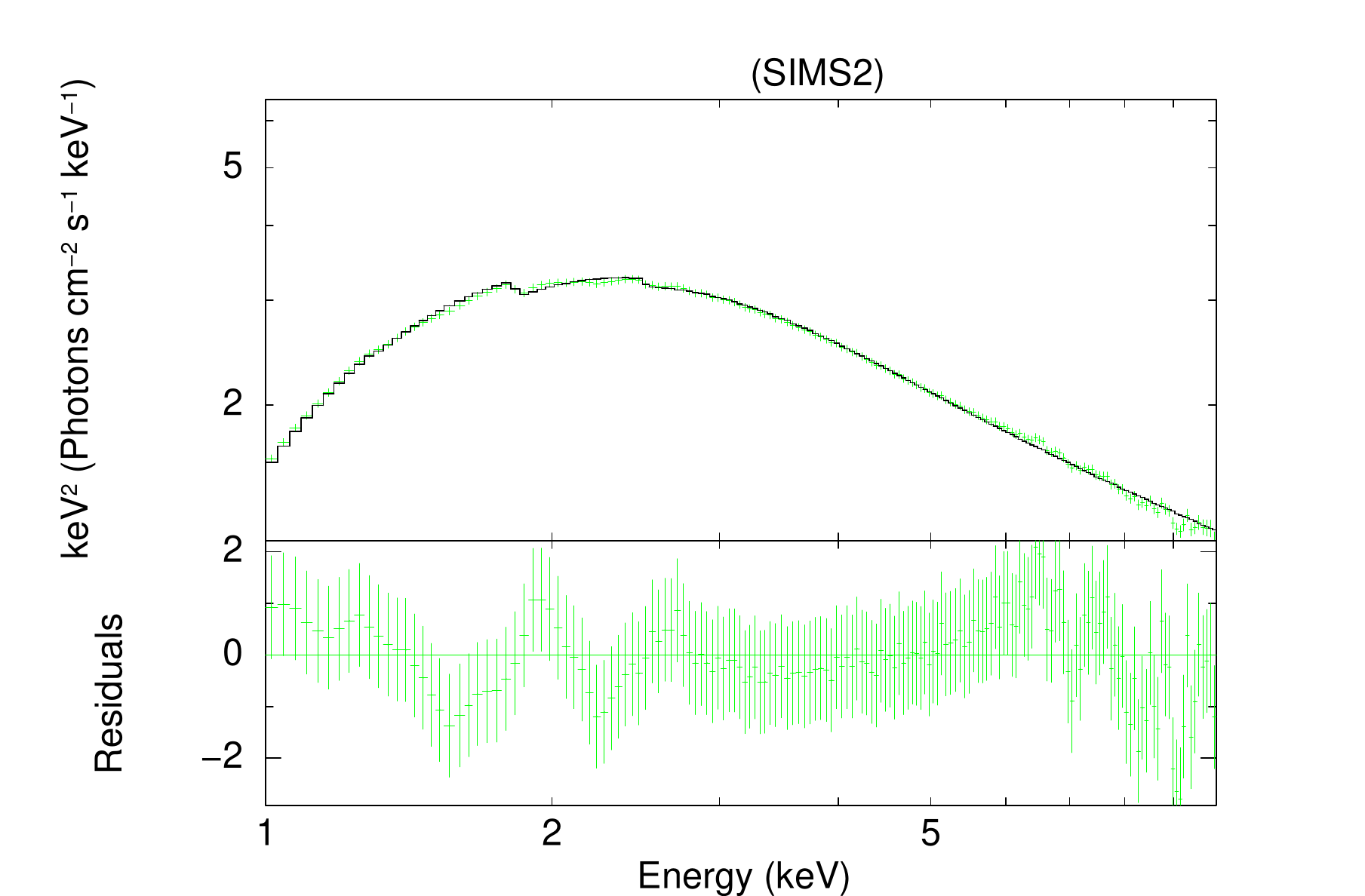}
    \includegraphics[width=5cm,angle=-90]{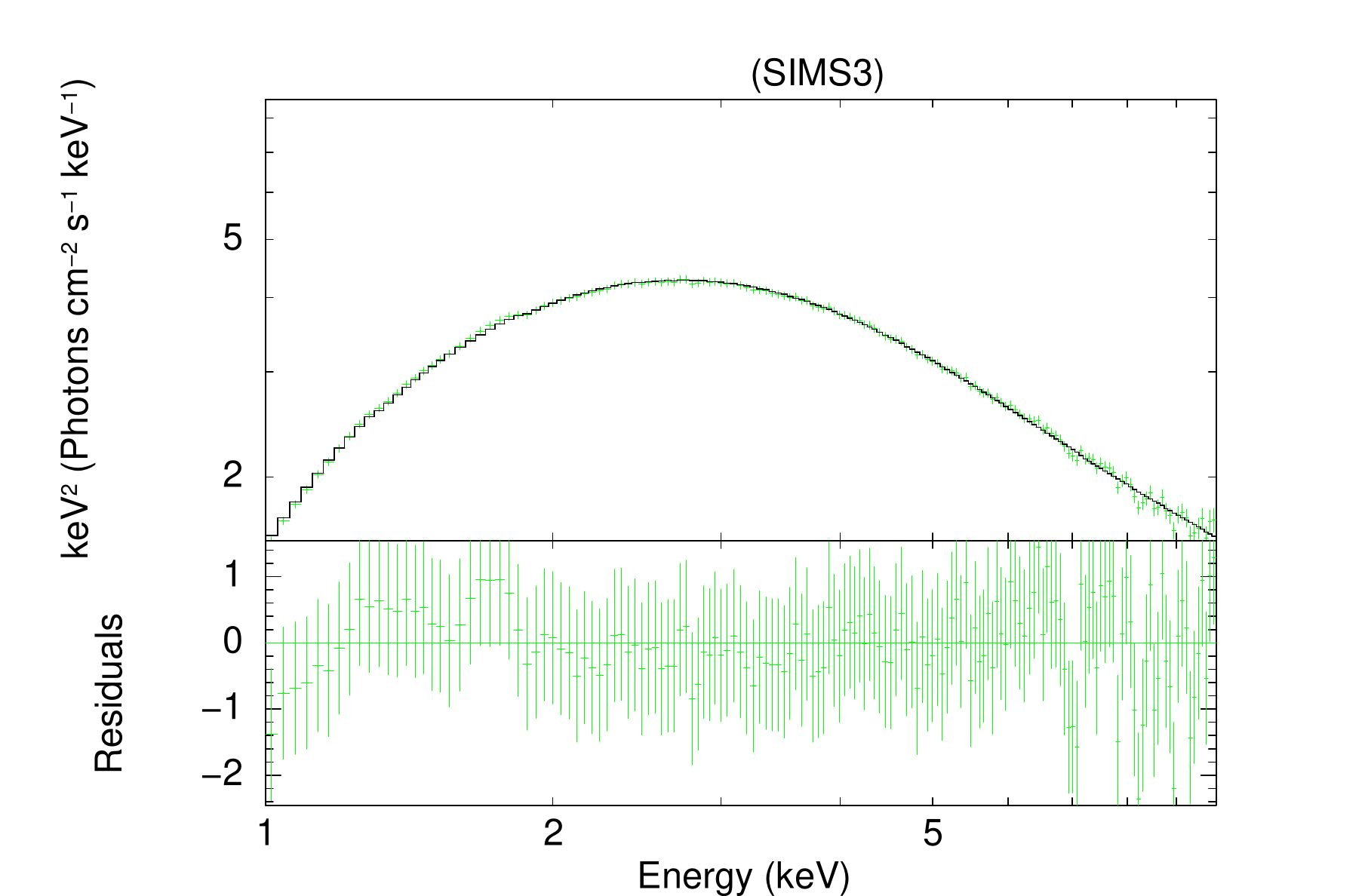}
    \includegraphics[width=5cm,angle=-90]{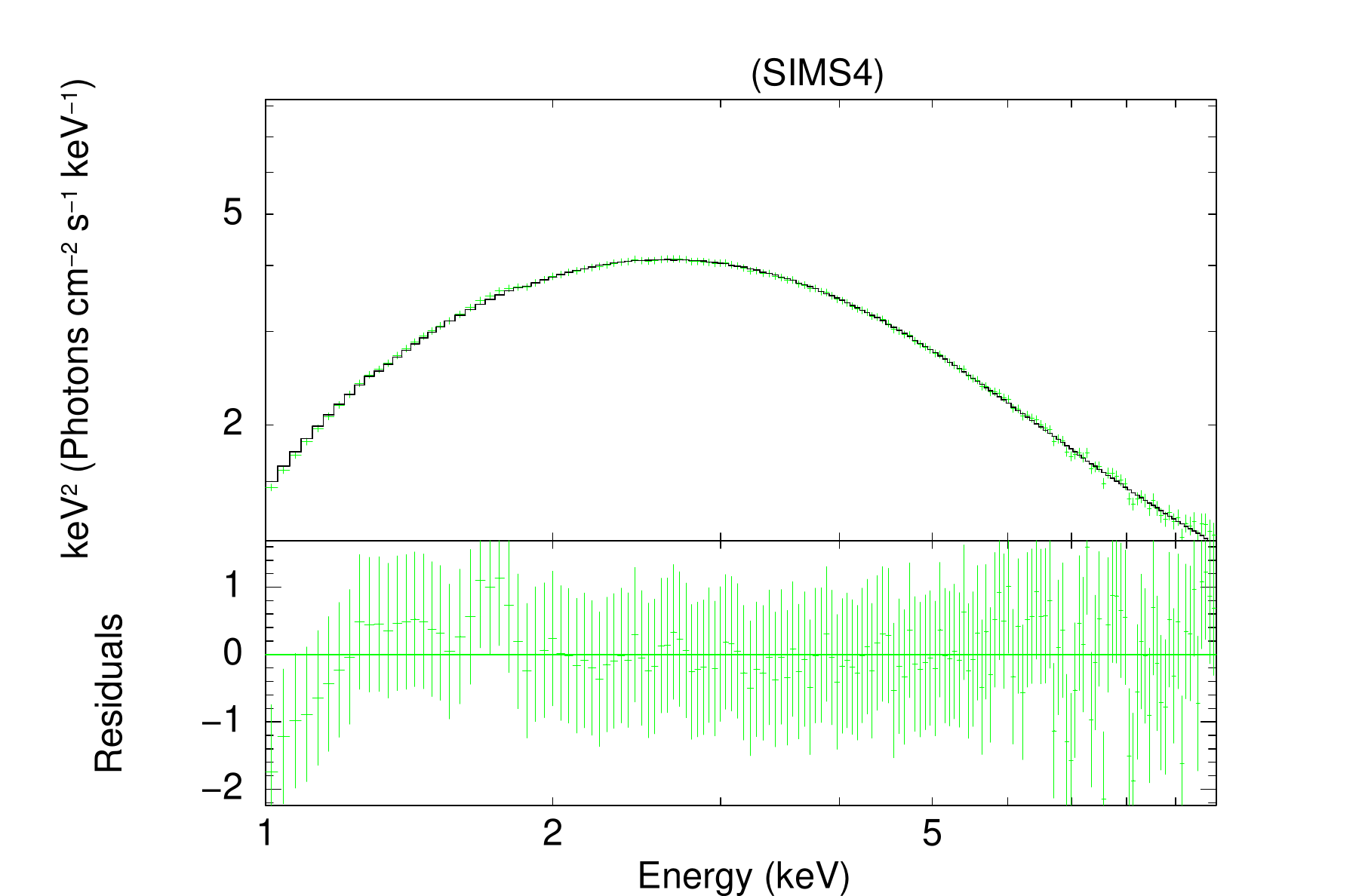}
        \includegraphics[width=5cm,angle=-90]{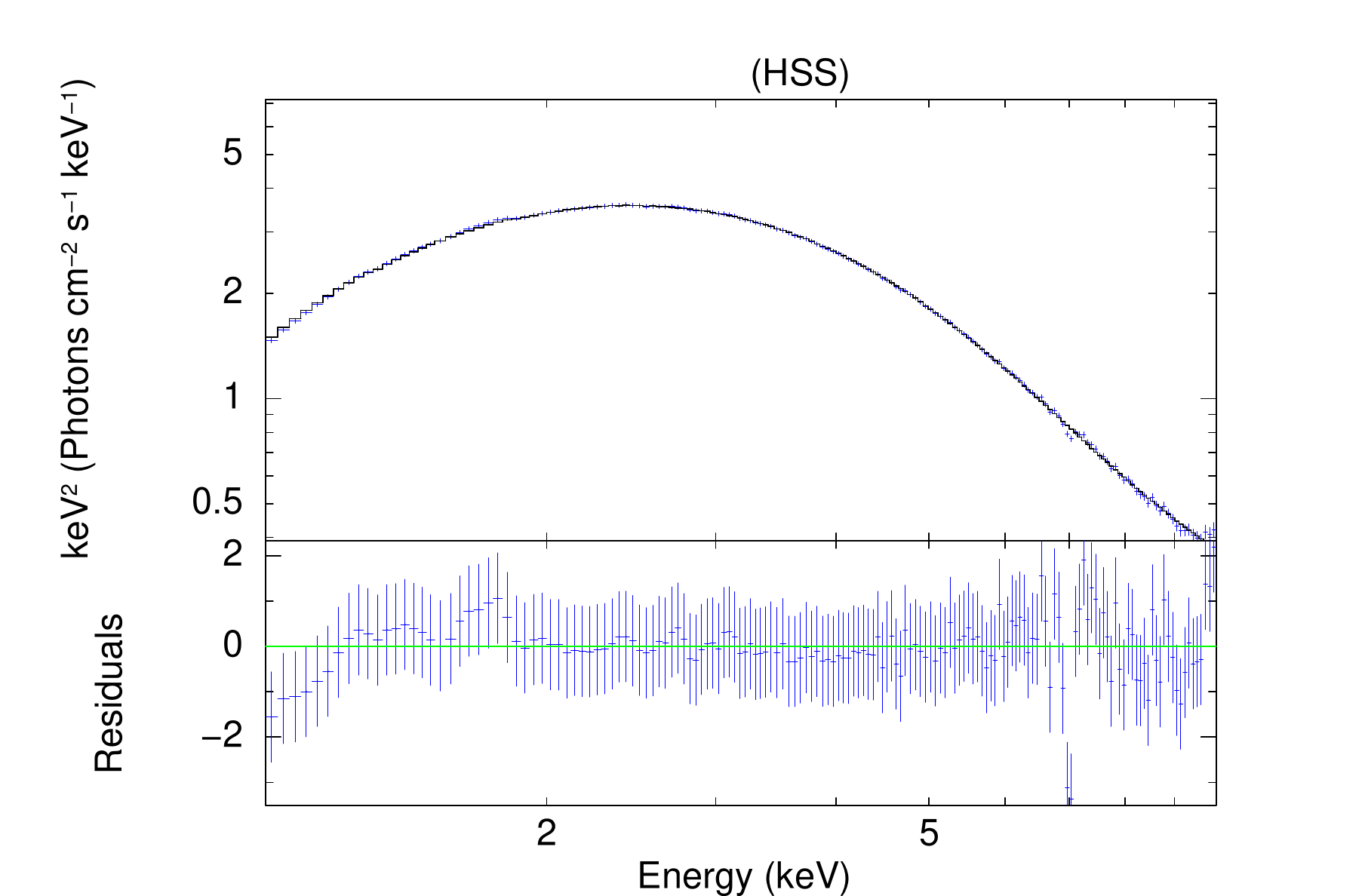}
    \caption{Photon spectra of different spectral states. \textit{NICER} segments HS, SIMS1, SIMS2, SIMS3, SIMS4, and HSS are in the 1-10 keV energy range, while \textit{AstroSat}/LAXPC (HIMS) is in the 3-30 keV energy range. The model used is $\texttt{constant}\times \texttt{Tbabs} \times \texttt{ThComp} \times \texttt{diskbb}$, where $kT_e$ is fixed at 100 keV. For HS and HIMS, we have added a \texttt{Gaussian} component fixed at 6.5 keV.}
    \label{fig:SPECTRA}
\end{figure*}

\begin{figure}[!ht]
    \centering
    \includegraphics[width=0.55\textwidth,keepaspectratio=true]{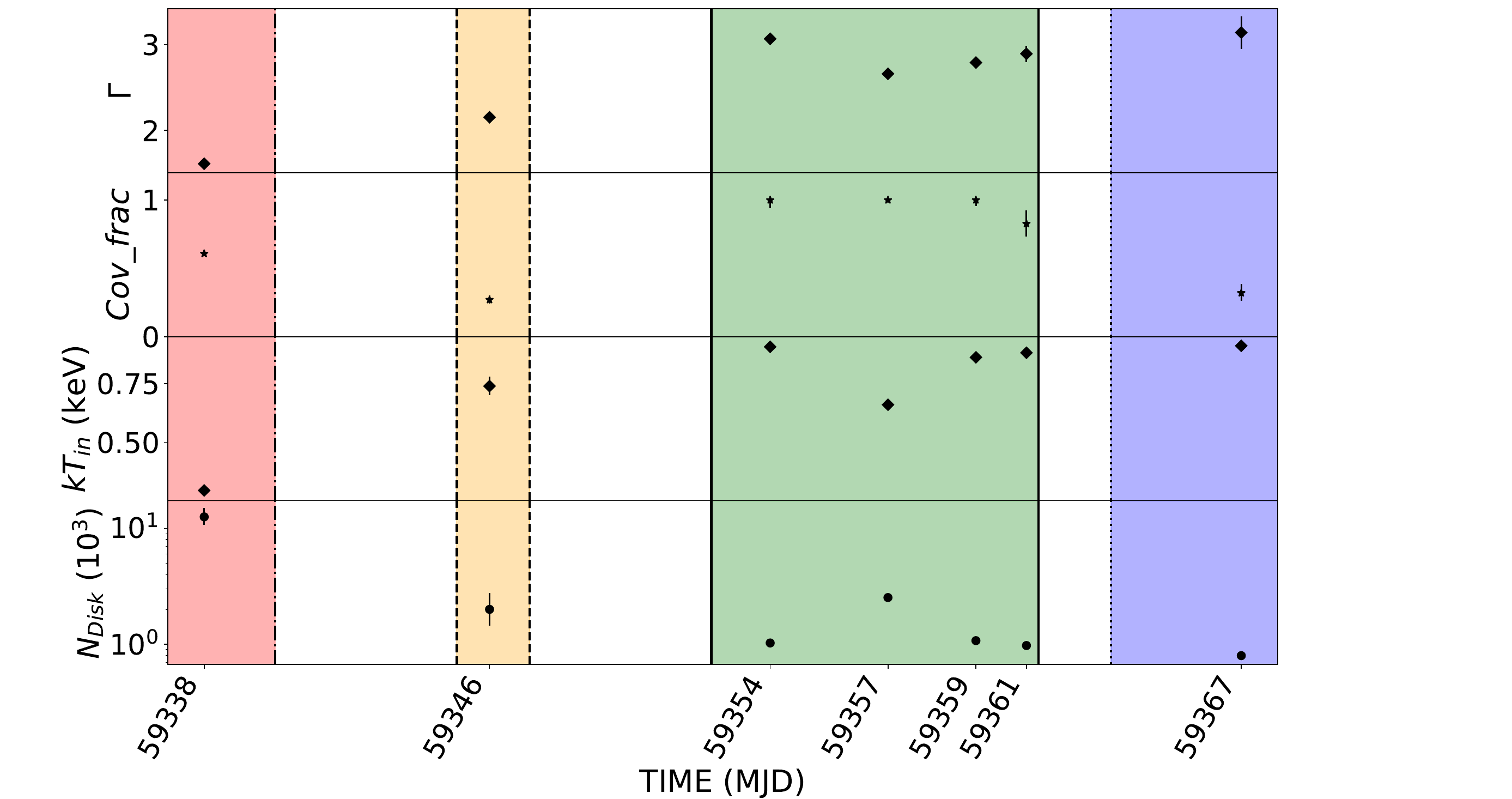}
    \caption{The evolution of spectral parameters with time (MJD). The errors are calculated at the 90\% confidence level. The MJDs of various segments were calculated as the mean of the starting and ending timings of distinct observations.} 
    \label{fig:paraVSseg}
\end{figure}
 
\subsection{Temporal Analysis}{\label{subsec:QPO}}
\subsubsection{The Power Density Spectrum (PDS)}
We started our analysis with the HIMS observation. We computed PDS for HIMS in a broad frequency range of 0.005-20 Hz and an energy band of 3-80 keV using \texttt{laxpc\_find\_freqlag} of LAXPC software. The PDS obtained are deadtime Poisson noise and background corrected \citep{2016ApJ...833...27Y}. Further, we fitted the PDS with a model consisting of multiple Lorentzians. As reported by \cite{2022ApJ...933...69C} and \cite{Jana_2022}, we detected Type-C QPOs in HIMS as shown in Figure \ref{fig:PDS_seg_1_7}.

Similarly, for \emph{NICER} observations, we analyzed different observations (HS, SIMS1, SIMS2, SIMS3, SIMS4 and HSS). We computed PDS in 0.005-20 Hz in an energy band of 0.2-12 keV for all \emph{NICER} segments using \texttt{nicer\_find\_freqlag} of \texttt{NICER\_RM\_Software}. We detected the presence of Type-B QPOs in SIMS1, SIMS2, and SIMS3 as shown in Figure \ref{fig:PDS_seg_1_7}, similar to that reported by \cite{2023MNRAS.523.4394Z} (taken during SIMS), and some aperiodic humps were observed in HS and SIMS4. 

Figure \ref{fig:PDS_seg_1_7} illustrates the PDS of all segments fitted with multiple Lorentzians. Table \ref{tab:A1} provides the details of all the Lorentzian components used to model the PDS of all segments. 


\subsection{Spectral Analysis}\label{subsec:SA}

We separately analyzed the \emph{AstroSat} and \emph{NICER} observations as they are not simultaneous and correspond to the different spectral states of the source. The time-averaged photon spectrum of XBs is often described using two spectral components: multicolor blackbody disk emission and thermal Comptonized emission from a hot corona. We used XSPEC components \texttt{diskbb} \citep{mitsuda1984energy} and \texttt{ThComp} \citep{zdziarski2020spectral} to model the disk and coronal emissions of both the \emph{AstroSat} and \emph{NICER}  spectra. \texttt{diskbb} has two spectral parameters: inner disk temperature $(kT_{in})$ and normalization $(N_{Disk})$ whereas \texttt{ThComp} is a convolution model with four parameters, namely spectral photon index (or optical depth) $(\Gamma_{tau})$, electron temperature $(kT_e)$, covering fraction $(cov\_frac)$, and redshift $(z)$.

We started by fitting the LAXPC spectra of HIMS segment. We restricted the fitting to 3-30 keV by ignoring energies beyond 30 keV because of LAXPC background uncertainties. We accounted for absorption column density $(N_H)$ using the XSPEC model \texttt{Tbabs} \citep{wijnands1999complex}. We first modeled the joint spectrum for HIMS with \texttt{Tbabs$\times$ThComp$\times$diskbb} and obtained a reduced $\chi^2$ of 121.85/24 $\sim$ 5. The systematic error of 2\% is incorporated in the analysis \citep{2017ApJ...Ranjeev}. It is to be noted that as \texttt{ThComp} model is a convolution model, we extended the energy range by applying ``\texttt{energies 0.01 500 1000 log}". 

We further incorporated a \texttt{Gaussian} component with line energy fixed at 6.5 keV and found that the reduced $\chi^2$ improved to $\sim$ 19.52/22 $\sim$ 0.9 for a width of $\sim$ 1.1 $\pm$ 0.1 keV. Moreover, the absorption column density N$_{H}$ turns out to be $\sim$ 0.23$\times$ 10$^{22}$ cm$^{-2}$ which is similar to the value reported in \citet{zhang2024variable}. Thus, the final model used to fit the \emph{AstroSat}/LAXPC spectrum is $\texttt{Tbabs} \times (\texttt{ThComp} \times \texttt{diskbb} + \texttt{Gaussian})$, with $kT_e$ fixed at 100 keV. 


Next, we fitted the 1-10 keV \emph{NICER} spectra with the model combination of  $\texttt{Tbabs} \times \texttt{ThComp}\times \texttt{diskbb}$. During fitting, we froze $N_H$ and ${kT_e}$ at the same value for all segments as for the \emph{AstroSat}/LAXPC case. For HS, this model combination gave a reduced chi-square of 1.47 (207.30/141). We observed a hump in residuals around $\sim$6.5 keV, and thereby, we added a \texttt{Gaussian} component fixed at 6.5 keV to resolve it, and we obtained the reduced chi-square 0.59 (80.88/139). 

For the SIMS1 segment, we fitted the spectrum using $\texttt{Tbabs} \times \texttt{ThComp} \times \texttt{diskbb}$ and obtained a reduced $\chi^2$ of 0.3 (44.64/163). Based on this, we adopted this model as the best-fit configuration and applied it to the remaining segments. It is important to note that the \texttt{Gaussian} component was used only for the HIMS (LAXPC) and HS (\emph{NICER}) segments, where significant residuals around 6.5 keV were observed. All other segments were fitted without a Gaussian component. Using the default systematic error from \texttt{nicerl3-spect}, we obtained reduced $\chi^2$ values in the range of 0.3–0.6 across all \emph{NICER} observations (HS, SIMS1, SIMS2, SIMS3, SIMS4, and HSS). These low reduced $\chi^2$ values for the \textit{NICER} segments likely indicate an overestimation of the default systematics. Figure \ref{fig:SPECTRA} shows the spectrum of all the \textit{NICER} segments fitted using the best-fit model. The residuals observed at energies less than and around 2 keV in Figure \ref{fig:SPECTRA} (\textit{NICER} segments) could be due to the calibration uncertainty at the Si and Au edges \citep{2021MNRAS.505.1213R}. In the SIMS2 segment, these instrumental features were more prominent in the residuals, and thus, we modeled them using two \texttt{edge} components at 1.8 keV and 2.5 keV.

Table \ref{tab:Best-fit parameters} lists the values of best-fit spectral parameters, and Figure \ref{fig:paraVSseg} shows the evolution of spectral parameters with time for all the \emph{NICER} and \emph{AstroSat} segments (HS, HIMS, HSS, and the four SIMS segments). We have marked each parameter value of different states following the same color code as discussed before. 

\begin{table*}[htbp]
	\centering
	\hspace{-0.9cm}
	\small
	\setlength{\tabcolsep}{3pt}
	\renewcommand{\arraystretch}{1.9}
	\caption{The best-fit spectral parameters and reduced $\chi^2$ of different segments with errors calculated at the 90\% confidence level. The value of ${kT_e}$ is fixed at 100.0 keV for all segments.}
	\label{tab:Best-fit parameters}
	\begin{tabular}{cccccc}
		\hline \hline
		SEG & $\Gamma$ & $cov\_frac$ & $kT_{in}$ (keV) & $N_{Disk}$ & $\chi^2/\mathrm{dof}$ \\
		\hline \hline
		HS     & $ 1.61^{+0.02}_{-0.02}$ & $ 0.61^{+0.02}_{-0.02}$ & $ 0.29^{+0.01}_{-0.01}$ & $ 12600^{+1900}_{-2400}$ & $ 81.88/139$ \\
		HIMS   & $ 2.15^{+0.02}_{-0.02}$ & $ 0.27^{+0.03}_{-0.03}$ & $ 0.74^{+0.04}_{-0.04}$ & $ 2000^{+550}_{-780}$ & $ 19.60/22$ \\
		SIMS1  & $ 3.06^{+0.01}_{-0.05}$ & $ >0.94$                & $ 0.91^{+0.01}_{-0.01}$ & $ 1030^{+20}_{-33}$ & $ 44.64/163$ \\
		SIMS2  & $ 2.66^{+0.01}_{-0.01}$ & $ >0.99$                & $ 0.66^{+0.01}_{-0.01}$ & $ 2530^{+60}_{-60}$ & $ 99.86/164$ \\
		SIMS3  & $ 2.79^{+0.01}_{-0.03}$ & $ >0.96$                & $ 0.86^{+0.01}_{-0.01}$ & $ 1080^{+30}_{-30}$ & $ 63.77/155$ \\
		SIMS4  & $ 2.9^{+0.1}_{-0.1}$    & $ 0.8^{+0.1}_{-0.1}$    & $ 0.88^{+0.01}_{-0.01}$ & $ 980^{+50}_{-50}$  & $ 57.93/155$ \\
		HSS    & $ 3.1^{+0.2}_{-0.2}$    & $ 0.32^{+0.07}_{-0.06}$ & $ 0.91^{+0.01}_{-0.01}$ & $ 800^{+20}_{-20}$  & $ 73.57/159$ \\
		\hline \hline
	\end{tabular}
\end{table*}


\subsection{Extraction of rms and lag}\label{section:RMSLAG}
\begin{figure*}[!ht]
    \centering
        \includegraphics[width=0.45\textwidth,keepaspectratio=true]{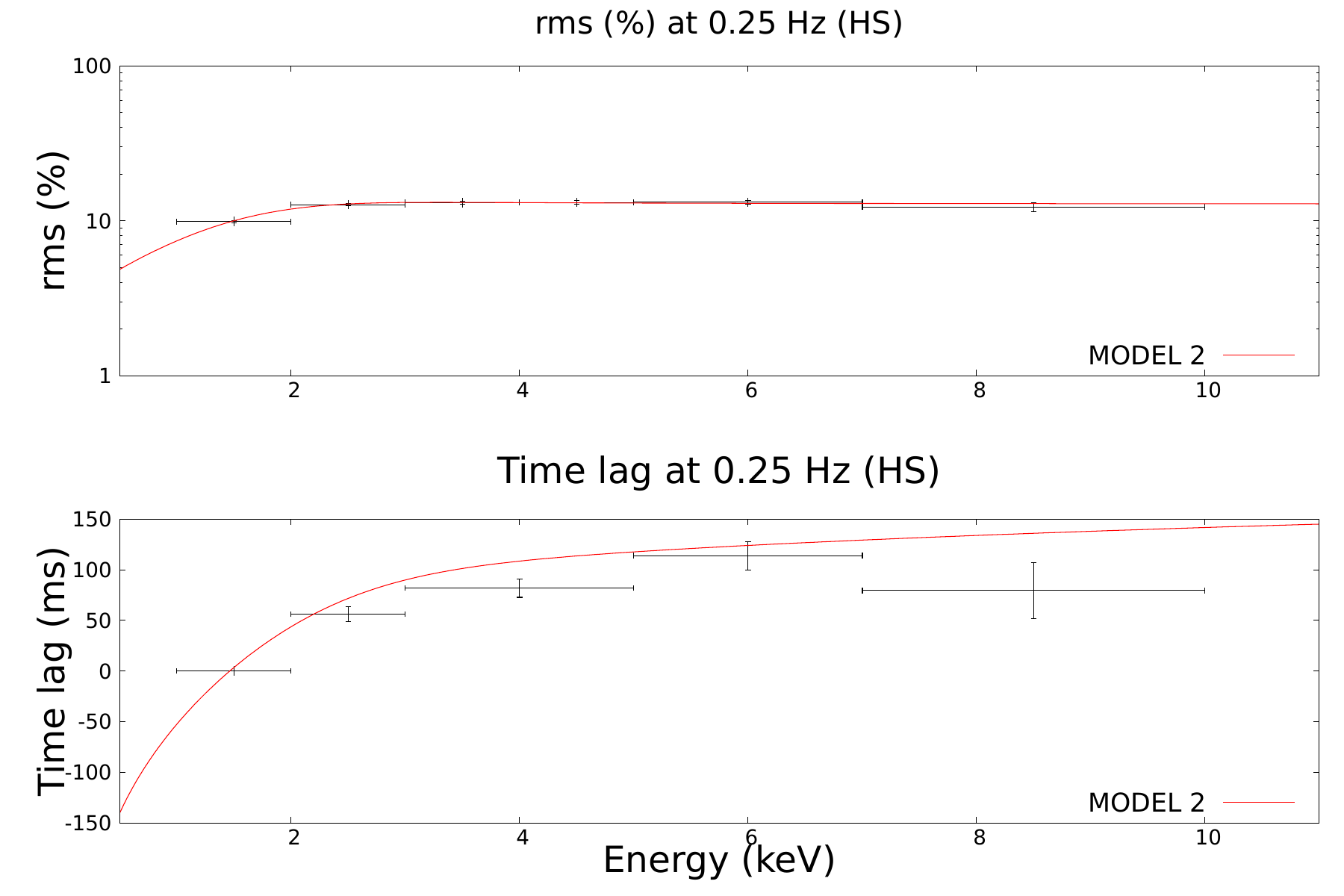}
    \includegraphics[width=0.45\textwidth,keepaspectratio=true]{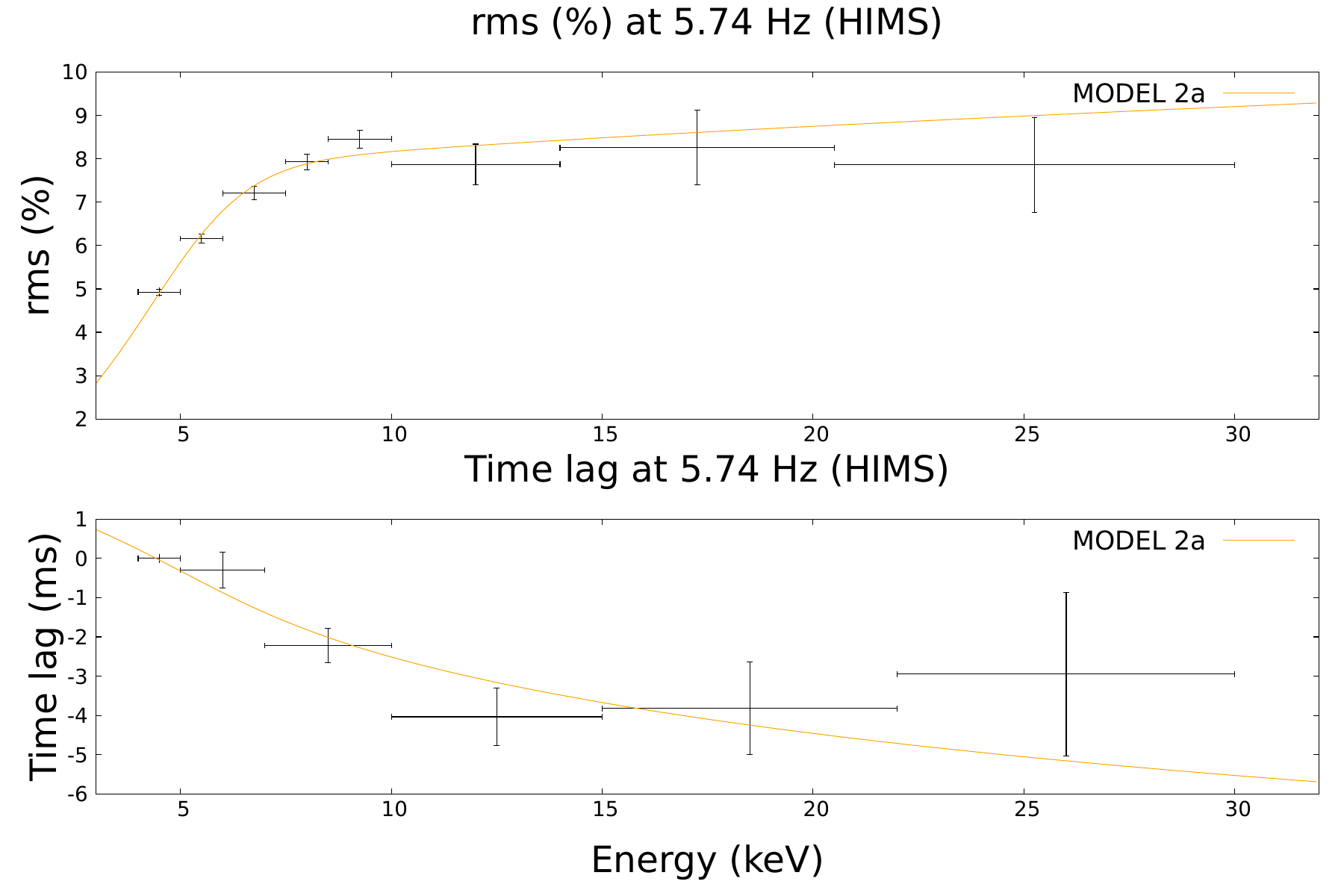}
        \includegraphics[width=0.45\textwidth,keepaspectratio=true]{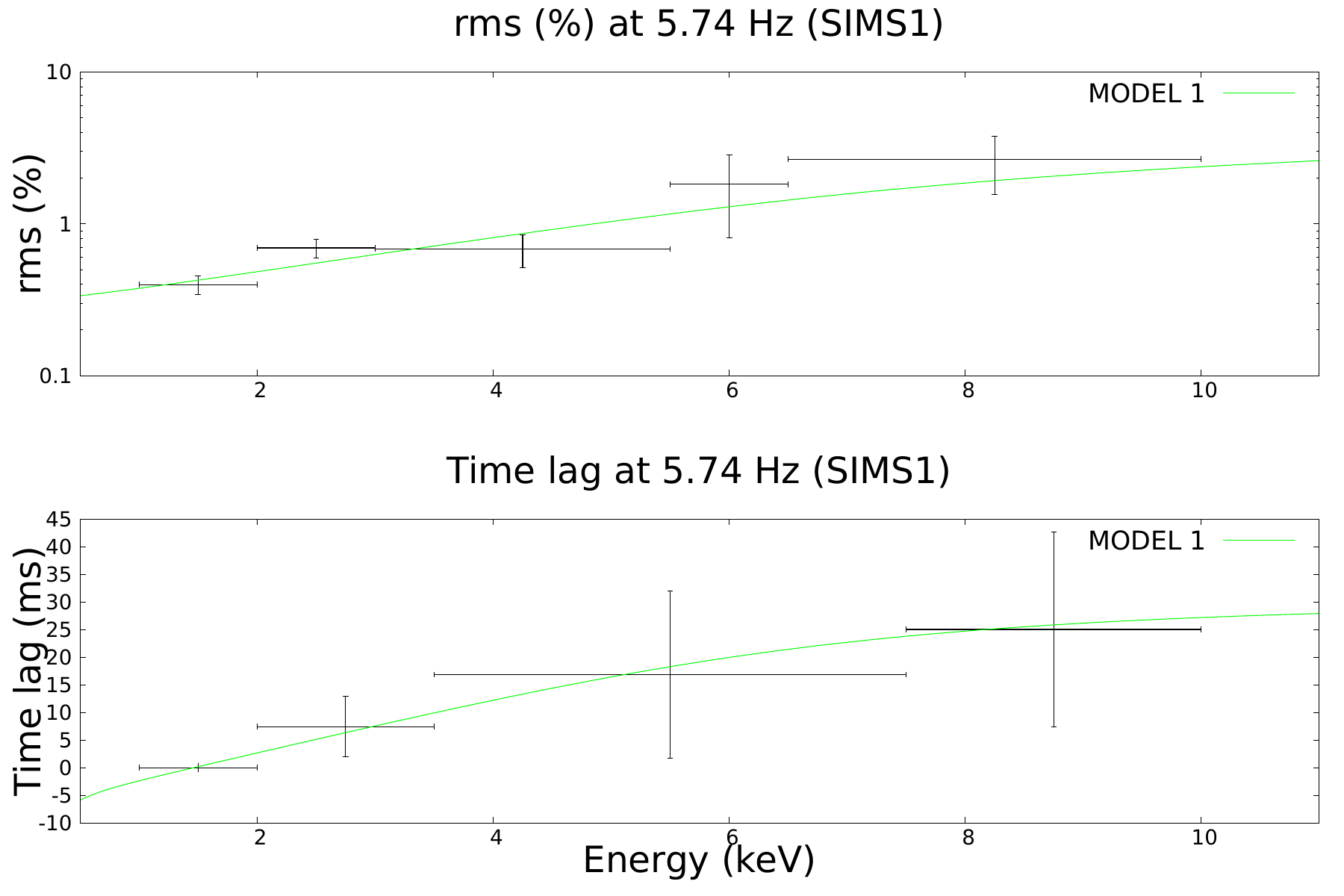}
    \includegraphics[width=0.45\textwidth,keepaspectratio=true]{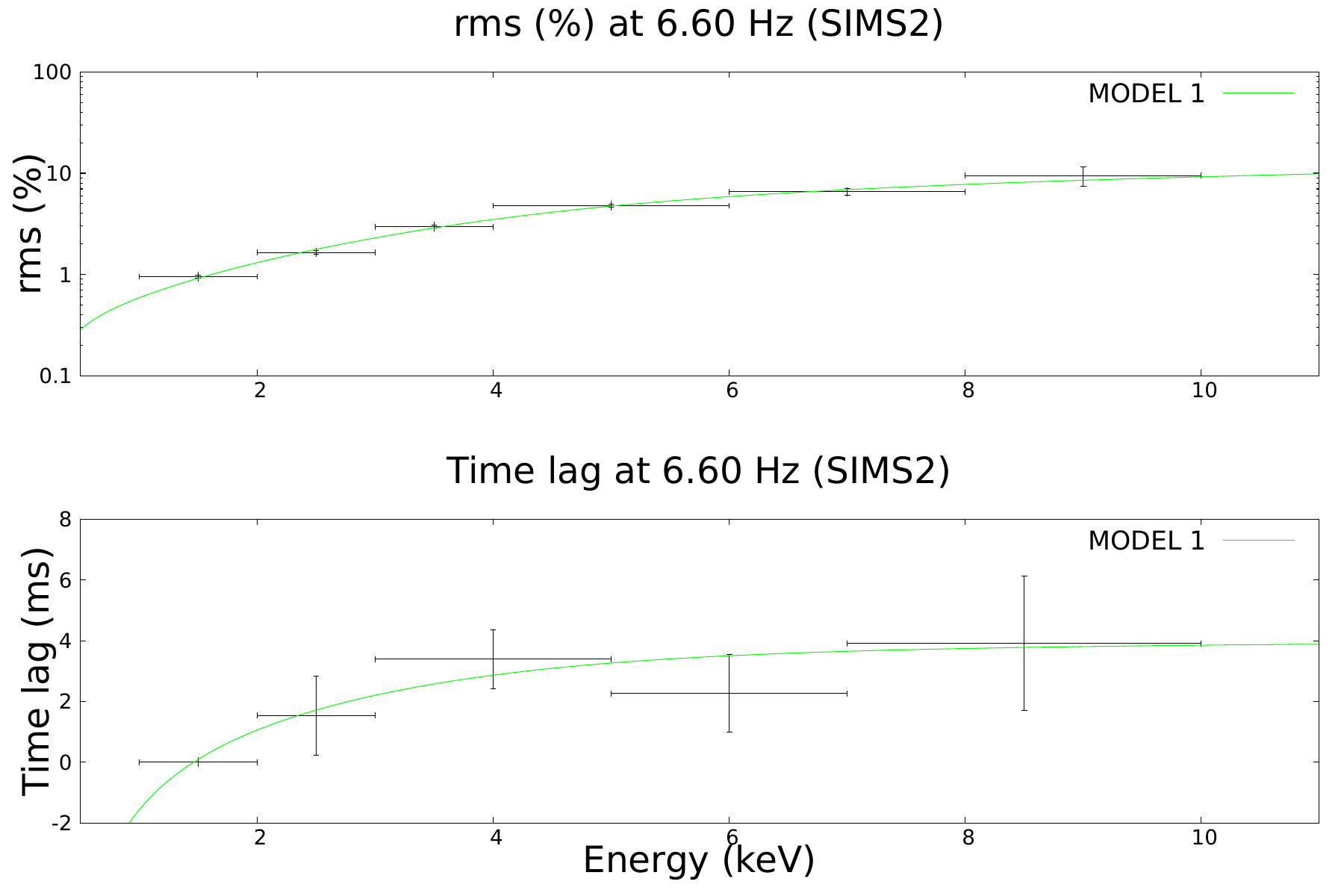}
    \includegraphics[width=0.45\textwidth,keepaspectratio=true]{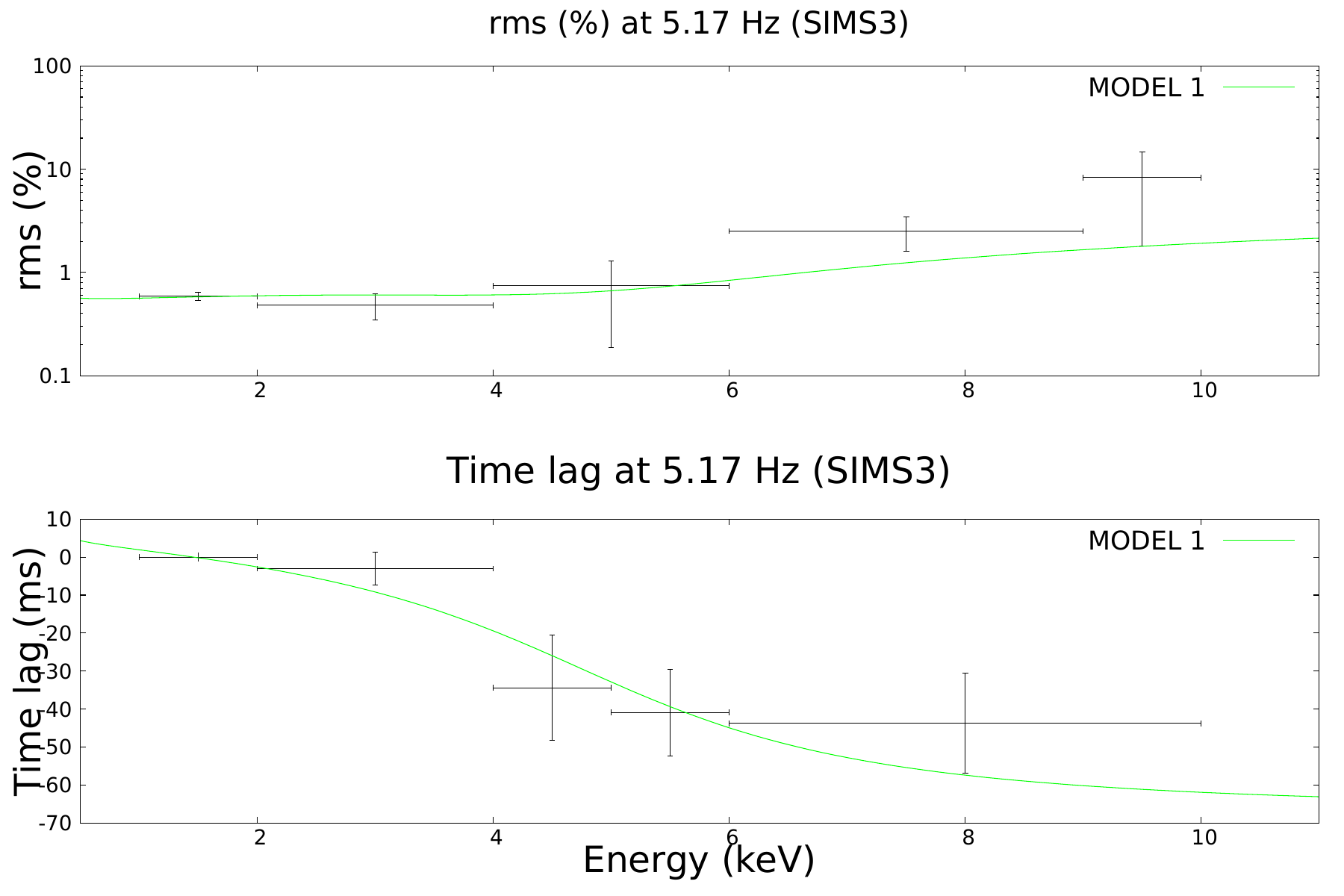}
        \includegraphics[width=0.45\textwidth,keepaspectratio=true]{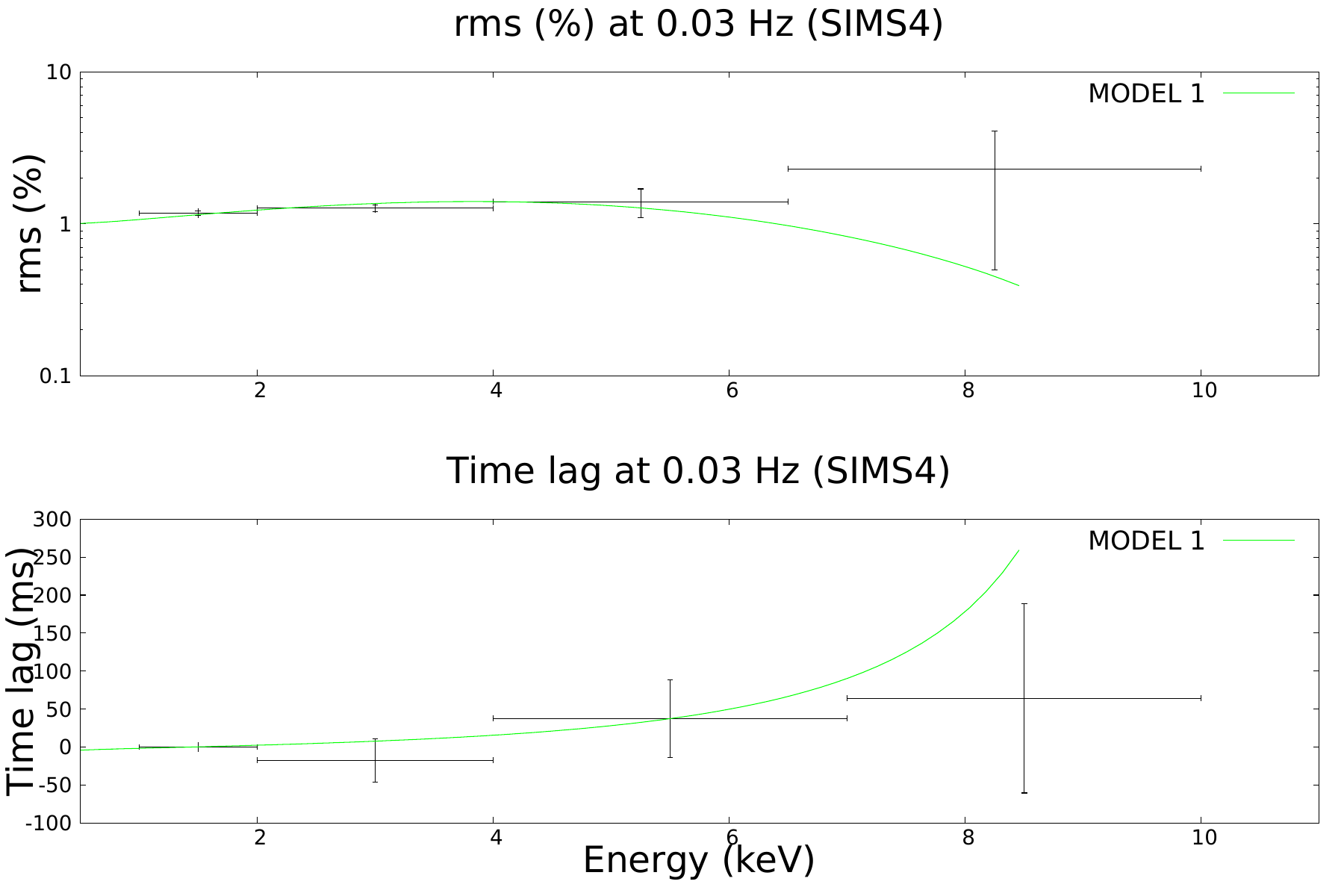}
    \caption{The observed rms and time lag in various segments fitted with different models. The plots illustrate different segment number, along with the peak frequency at which rms and time lag are determined. The fitted model for the observed rms and lag are also indicated within the plot. The parameter values of the model are given in Table \ref{tab:Physical_parameter_NICER}.}
    \label{fig:model_rms_lag}
\end{figure*}

We computed the rms and time lag energy spectra in 4-30 keV using the subroutine \texttt{laxpc\_find\_freqlag} at the centroid frequency of the detected QPO for \emph{AstroSat} observation (HIMS) as shown in Figure \ref{fig:model_rms_lag}. Similarly, we computed the rms and lag spectra in 1-10 keV using \texttt{nicer\_find\_freqlag} for all four SIMS segments as shown in Figure \ref{fig:model_rms_lag}.  

Both the subroutine estimates the rms and lag for different energy bins in particular energy range (for \emph{AstroSat} (4-30 keV) and for \emph{NICER} (1-10 keV)) using inputs like frequency resolution $\Delta$f (equal to or half of FWHM (full width at half maximum) of QPO frequency) and QPO frequency or the characteristic frequency of the Lorentzian ($\nu$ = $\sqrt{{\nu_{0}}^{2}+ \Delta^{2}}$, where $\nu_{0}$ corresponds to centroid frequency and $\Delta$ is the half-width at half-maximum (HWHM=$\sigma/2$)). The rms is calculated as the square root of the integration of the power spectra in the frequency range f - $\Delta$f and f + $\Delta$f for a particular band. Along with the rms, phase lag is also computed by the subroutine for the cross-spectra of the two light curves of different energy bands using one as a reference energy band (E$_{ref}$). This subroutine also uses different inputs, such as the minimum frequency and maximum frequency, or the Nyquist frequency.

 We observed that the rms increased with energy for all the segments. However, we found different strengths of fractional variability in different states. In LHS (HS segment), rms for the broad feature is maximum at $\sim$13\% in 5-7 keV energy band with a positive time lag, which increased to 114 ms in 5-7 keV but then decreased to 80 ms in 7-10 keV energy band as it can be seen in Figure \ref{fig:model_rms_lag} (HS). The average fractional variability in the HIMS segment is $\sim$6\% in the 4-30 keV band. The average fractional variability in different segments of SIMS is between 2\% and 23\% over the full energy band (1-10 keV). The HIMS segment showed negative lags (soft lags), which decreased monotonically over the full energy range till $\sim$ 5 ms Figure \ref{fig:model_rms_lag}. We observed that the nature of the time lags shifted to negative in SIMS3 with a maximum negative lag of 50 ms for the energy band 6-10 keV, while for SIMS1, SIMS2, and SIMS4, we observed the positive nature of the lag with the average positive lag in the range between $\sim$ 5 ms and $\sim$ 66 ms, as shown in Figure \ref{fig:model_rms_lag}. We found insignificant variations in the fractional variability and time lag during the HS segment. 

\subsection{Modelling the rms and time lag spectra}\label{section:modelling}
Next, we modeled the energy-dependent rms and time lag computed at different frequencies of QPOs and aperiodic bumps detected in HS, HIMS, and the four SIMS segments. We used the technique given by \cite{garg2020identifying}, which models the energy-dependent temporal properties using the spectral information of the source. This model considers an accretion flow consisting of a geometrically thin and optically thick disk truncated at some radius far from ISCO and a hot, optically thin, and geometrically thick corona consisting of high-energy electrons near the BH. \citet{garg2020identifying} models both outer truncated disk and coronal emissions using the XSPEC components \texttt{diskbb} and \texttt{ThComp}, respectively. In their model, they also translated the spectral parameter $kT_e$  into a more physical parameter, heating rate ($\dot{H})$ for the analysis. So, there are five parameters $kT_{in}$, $\tau$, $\dot{H}$, $f_{sc}$ (equivalent to parameter $cov\_frac$), and $N_{disk}$ which are obtained by fitting the time-averaged photon spectrum.
According to \citet{garg2020identifying}, the variations on the spectral parameters can cause the variation in the steady state spectrum ( $F(E)$) given as
	\begin{equation}
		\Delta F (E) = \sum_{j=1}^M \frac{\partial F (E)}{\partial \alpha_j} \Delta \alpha_j
		\label{DeltaSE}
	\end{equation}
	\noindent
	where $\alpha_j$ are the spectral parameters and the small variations of these parameter values are denoted by $\Delta \alpha_j$, and $M$ is the number of parameters. $\Delta \alpha_j$ can be written as $A_j exp^{i\omega_jt}$, $A_j$ is the amplitude, and $\omega_j$ denotes the frequency of sinusoidal variations. They estimated rms using $(\frac{1}{\sqrt{2}})|\Delta F(E)|/F(E)$ and phase lag using the argument of $\Delta F(E_{ref} )\times \Delta F(E)$, where $E_{ref}$ is the reference energy band. The phase lag can be easily calculated as the product of the time lag and 2$\pi$$\nu$ where $\nu$ is the frequency of the signal. 

	The heating rate ($\dot H$) of the corona can be defined as the energy difference between the emitted Comptonized spectrum and the incoming seed photon flux. Mathematically, it can be defined by equation \ref{eq:5}.  
	
	\begin{equation}
		\dot{H} = \int
		E(F_{c}(E,kT_e,\tau)-F_{inp}(E)) dE
		\label{eq:5}
	\end{equation}
	\noindent
	where $F_c$ is the photon flux from the corona and $F_{inp} (E) = f*F_{d}$ is the seed photon flux. \\ 
		  The predicted rms and lag can then be compared with the observed energy-dependent rms and lag spectra. Further, \cite{garg2022energy} converted the parameters $\delta N_{disk}$ and $\delta kT_{in}$ into $\delta R_{in}$ and $\delta \dot{M}$ respectively, using the relations $R_{in}$$\propto$ $N_{disk}^{1/2}$ and $\dot{M}\propto T_{in}^4 R_{in}^3$. Thereby, they fitted the rms and lag spectra using the variations in $\dot{M}$, $\tau$, $\dot{H}$, $f_{sc}$, and $R_{in}$.
	

\begin{table*}[htbp]
	\caption{The physical parameters of various models in different segments where errors are calculated at 1$\sigma$ level.}
	\label{tab:Physical_parameter_NICER} 
	\hspace{-1.0cm}
	\centering
	\small
	\setlength{\tabcolsep}{2pt} 
	\renewcommand{\arraystretch}{1.2} 
	\begin{tabular}{@{}ccccccccccc@{}}
		\hline
		Frequency (Hz) & SEG & Model & $|\delta \dot{M}|$ (\%) & $|\delta \dot{H}|$ (\%) & $\phi_{\dot{H}}$ & $ T_{\dot{H}}$ (ms) & $|\delta f_{sc}|$ (\%) & $\phi_{f_{sc}}$ & $ T_{f_{sc}}$ (ms) & $\chi^2$/dof \\
		\hline\hline
		$ 0.25$ & HS    & MODEL 2 & $ 9.3^{+0.9}_{-0.9}$     & $ 15.6^{+3.9}_{-3.9}$     & $ 0.4^{+0.2}_{-0.2}$ & $ 264^{+120}_{-110}$ & $ 9.9^{+2.1}_{-2.1}$   & $ 0.47^{+0.22}_{-0.22}$ & $ 300^{+140}_{-140}$ & $ 3.30/5$ \\
		$ 5.74$ & HIMS  & MODEL 2a & $ 1.9^{+0.2}_{-0.2}$     & $ 12.7^{+0.6}_{-0.6}$     & $ -0.20^{+0.03}_{-0.03}$ & $ -5.6^{+0.7}_{-0.7}$  & $ 8.6^{+1.0}_{-1.0}$   & \multicolumn{1}{c}{--} & \multicolumn{1}{c}{--} & $ 11.42/9$ \\
		$ 5.74$ & SIMS1 & MODEL 1 & $ 0.7^{+0.1}_{-0.1}$     & $ 3.4^{+1.1}_{-1.1}$      & $ 1^{+0.4}_{-0.4}$ & $ 28^{+12}_{-11}$ & \multicolumn{1}{c}{--} & \multicolumn{1}{c}{--} & \multicolumn{1}{c}{--} & $ 4.43/5$ \\
		$ 6.60$ & SIMS2 & MODEL 1 & $ 1.25^{+0.05}_{-0.05}$  & $ 10.9^{+0.3}_{-0.3}$     & $ 0.17^{+0.04}_{-0.04}$ & $ 4.2^{+1.0}_{-1.0}$ & \multicolumn{1}{c}{--} & \multicolumn{1}{c}{--} & \multicolumn{1}{c}{--} & $ 6.44/7$ \\
		$ 5.17$ & SIMS3 & MODEL 1 & $ 0.9^{+0.1}_{-0.1}$     & $ 4^{+1}_{-1}$      & $ -2.1^{+0.3}_{-0.4}$ & $ -64^{+10}_{-11}$ & \multicolumn{1}{c}{--} & \multicolumn{1}{c}{--} & \multicolumn{1}{c}{--} & $ 7.17/6$ \\
		$ 0.03$ & SIMS4 & MODEL 1 & $ 1.8^{+0.05}_{-0.05}$   & $ 0.9^{+0.7}_{-0.2}$      & $ <1.2$ & $ <6250$ & \multicolumn{1}{c}{--} & \multicolumn{1}{c}{--} & \multicolumn{1}{c}{--} & $ 4.44/4$ \\
		\hline \hline
	\end{tabular}
\end{table*}

First, we modeled the rms and lag spectra at Type-C QPO frequency detected in the HIMS segment. Following \citet{garg2022energy}, we first fitted a simple MODEL 1, which consisted of variations in the accretion rate ($\delta \dot{M}$) and the heating rate ($\delta \dot{H}$) along with a phase delay between them. The fitting gave a reduced $\chi^2$ of $>$ 2. Next, we introduced variation in $f_{sc}$ with some phase delay ($\phi_{f_{sc}}$) with respect to the accretion rate, and we obtained a good fit with a reduced $\chi^2$ of $\sim$ 1.43 (11.42/8). We call this model MODEL 2. 
However, we found that $\phi_{f_{sc}}$ was consistent with zero during the fitting, and therefore fixed it to zero, assuming no delay with respect to the accretion rate. By doing this, we obtained $\chi^2/dof$ as 11.42/9 $\sim 1.26$. We call this model MODEL 2a. The parameter values of the fitting of the MODEL 2a on HIMS are given in Table \ref{tab:Physical_parameter_NICER}. 


In Table \ref{tab:Physical_parameter_NICER}, the parameters expressed in percentage (\%) provide an intuitive measure of the relative modulation strength, allowing for a direct comparison of which parameters contribute most to the observed variability. The phase lag values are used to compute the corresponding time lags ($T_{\dot{H}}$ and $T_{f_{\text{sc}}}$) between parameter variations, expressed in milliseconds. Depending on the sign and value of the phase lag, as well as the percentage variations in the disk and coronal parameters, one can infer from which parameter the variability originated, the timescale and direction of its propagation, and which parameter began to vary in response to the perturbation. For instance, in the HIMS segment, the rms and lag spectra are fitted using MODEL 2a, which incorporates variations in the accretion rate, coronal heating rate, and scattering fraction, along with the lag w.r.t. the accretion rate. The results suggest that the coronal variations are due to the variations in both the heating rate and scattering fraction parameters. Moreover, the negative lags indicate that the coronal variations lead the accretion rate, implying an outward propagation of perturbations from the corona to the disk.

Further, we fitted the rms and lag spectra for Type-B QPOs detected in SIMS's segments (SIMS1, SIMS2 and SIMS3). We first fitted the rms and lag of SIMS1 with the same MODEL 1 as previously and obtained a reduced $\chi^2$ of $\sim$ 0.92 as shown in Figure \ref{fig:model_rms_lag}. Next, we found that reduced $\chi^2$ varies from 0.89 to 1.19 for SIMS1, SIMS2, and SIMS3 segments, indicating that MODEL 1 is a good fit for the observed rms and lag spectra. We refrained from further fitting with other models so as not to overfit the data. Further, we fitted the rms and lag spectra for the broad feature seen for SIMS4 with MODEL 1 and obtained a good fit with reduced $\chi^2$ of $\sim$ 1.11.  Table \ref{tab:Physical_parameter_NICER} lists all three best-fit parameters and reduced $\chi^2$ for all SIMS segments.

Lastly, we modeled the rms and time lag of the aperiodic hump observed in the power spectra of the HS segment. We found that MODEL 1 can't fit the observed rms and lags. So, we tried with MODEL 2 and got a fit of a reduced $\chi^2$ of $\sim$ 0.66. The parameter values of the fit are given in Table \ref{tab:Physical_parameter_NICER}. We restrain further variation in parameters to avoid overfitting. Lastly, we found insignificant variations for HSS in the soft state and refrained from modeling its rms and lag spectra.

\section{DISCUSSION AND CONCLUSIONS}\label{sec:DC}
 
We performed the spectral and timing study of MAXIJ1803 using \emph{AstroSat} and \emph{NICER} observations. We used the month-long observations of the source during its 2021 outburst as it transited through LHS, HIMS, SIMS, and HSS. It became quite favorable that \emph{NICER}, which works in 0.2-12 keV, captured the source during the SIMS and HSS, which are dominated predominantly by soft emission. Additionally, \emph{NICER} also observed the source during LHS, capturing the hard emission in soft energy bands. Similarly, \emph{AstroSat}/LAXPC, which works in 3-80 keV, observed the source during HIMS. We could perform the spectral and temporal analysis for all the spectral states of the source. 

The MAXI light curve, along with hardness variations (Figure \ref{fig:Sub-figure 1} (a) and (b)) and the HID plot (Figure \ref{fig:Sub-figure 1} (c)), show the rise and decline of the source through the outburst. Though the MAXI count rates show the expected behavior for all the states, hardness variations seem to drop from a high value for LHS to a smaller value, which stayed constant during HIMS, SIMS, and HSS. Further, the source trajectory in the HID doesn't completely resemble but is somewhat like a `Q' pattern as expected for BHXBs \citep{2005A&A...440..207B}. The radio observations taken from \citet{2023MNRAS.522...70W} are marked as stars in HID to distinguish among spectral states as provided by \citet{2023MNRAS.522...70W} (in Table 1). We further confirmed this using our spectral and temporal analysis.

Our spectral analysis revealed that the source emission in all the states can be described as the blackbody emission from an outer thermal disk and the Comptonized emission from the coronal region. Interestingly, the parameter evolutions obtained in our spectral states also suggest the spectral transitions across different spectral states. Moreover, we found a low inner disk temperature and a high disk normalization, which indicates the presence of a cold disk being truncated at a large distance from the BH and coronal regions dominating the X-ray emission, which is a typical feature of a low-hard state \citep{1995Narayan,2007Liu}.


We also detected a Gaussian line emission in the HS segment, peaking around $\sim$ 6.5 keV with $\sigma \sim 0.9^{+0.2}_{-0.2}$ and norm $\sim$ $0.005^{+0.002}_{-0.001}$. The presence of narrow or broad reflection line components in the hard state spectra has previously been reported in sources like Cyg X-1 \citep{2015ApJParker} and MAXI J1820+070 \citep{2019Natur.565..198K,2019MNRAS.490.1350B}. The HS often consists of compact jets with a spectrum ranging from radio wavelength to near-IR wavelength \citep{2011ApJ...740L..13G}, which likely extends to X-rays and $\gamma$-rays. The compact jets can emit X-rays from different heights, illuminating both the inside and outside of the truncated disk, leading to reflection features in the hard state \citep{2019Natur.565..198K,2019MNRAS.490.1350B}.
 
\begin{table*}[!ht]
\centering
    \caption{The table gives information about the inferences made by modeling the rms and time lag of different variability features observed during different spectral states.}
    \label{tab:my_label}
 \begin{tabular}{ccccc}
    \hline
    \textbf{Spectral states} & \textbf{Type of } & \textbf{Nature of lag} &  \textbf{Initial Variation} & \textbf{Variation after} \\
    & \textbf{variability feature} & &  &\textbf{ a time delay} \\
    \hline
       LHS  & Broad feature & Positive lag  & Accretion rate & Coronal heating rate\\
     (HS segment)  & & & &  and fractional scattering \\ \\
        HIMS  & Type-C QPO& Negative lag & Coronal heating rate  & Accretion rate  \\
        (HIMS segment) & & & & and fractional scattering\\\\
        SIMS & Type-B QPO & Positive lag & Accretion rate & Coronal heating rate  \\
         (SIMS1 and SIMS2 segment) & & & & \\\\
        SIMS  & Type-B QPO & Negative lag & Coronal heating rate  & Accretion rate  \\
        (SIMS3 segment) & & & & \\\\
        SIMS  & Broad feature & Positive lag & Accretion rate & Coronal heating rate \\
        (SIMS4 segment) & & & & \\\\
        HSS  &- &- &- &- \\
        (HSS segment)\\
        \hline
    \end{tabular}
\end{table*}
Next, as the source moved to the HIMS segment, there was a large decrease in the disk normalization and an increase in the inner disk temperature, indicating that the source transited to the HIMS. Further, as the source moves through the four SIMS segments, the $kT_{in}$ continues to increase with disk normalization becoming very small. Collectively, it shows that the truncated disk has advanced towards the BH, and thereby, the emitted flux is getting softer, which is often seen for SIMS and HSS \citep{2011BASI...39..409B}. Our results are consistent with the findings of \citet{2023MNRAS.522...70W}, which used radio observations as discussed in Section \ref{sec:NICER_OBS}. We also found an absorption feature in the residuals of our spectrum in all four SIMS segments and in HSS but we refrain from modeling it with \texttt{Gaussian}, as the reduced $\chi^2$ values are overestimated due to the default systematics of the \texttt{nicerl3-spect}. \citet{zhang2024variable} also reported such absorption features in MAXIJ1803 in different epochs of SIMS and HSS, indicating the presence of winds.

Our temporal analysis showed that the power spectrum for LHS possesses broad features, but as the source transited to HIMS, Type-C QPOs along with sub-harmonic appeared at the frequencies of $\sim$  5.75$\pm$0.08 Hz and 1.16$\pm$0.06 Hz respectively. Further, as the source shifted to SIMS, Type-C QPOs disappeared, with Type-B QPOs appearing in the PDS within the range $\sim$ 5.17-6.61 Hz. We confirmed the types of QPOs using the quality factor (Q: $\nu_{centroid}/FWHM$), rms (\%), and the detection of QPOs in different spectral states. The QPO, detected in the HIMS segment, have the Q factor of $\sim$ 5 and the rms (\%) of $\sim$ 12\%, indicating that they are type C. The Q-factor of $\sim$ 5-10, rms of $\sim$ 0.63-2.84\%, and detection in SIMS \citep{2005Belloni} using \emph{NICER} segments, indicated type-B QPOs. Similar values of Q and rms (\%) have been reported by \citet{2022ApJ...933...69C,Jana_2022,2023MNRAS.523.4394Z}. Lastly, in the HSS, the variability is quite low, with only broad features seen in the PDS. Further, we couldn't detect any sharp correlations between the QPO frequency and spectral parameters over the source evolution.

We modeled the observed behavior of energy-dependent fractional rms and time lag using the technique devised by \citet{garg2020identifying}. For the Type-C QPOs detected in the HIMS segment, we discovered that the dynamic origin of the QPO is from the corona, which propagated away from the BH, causing perturbations in $f_{sc}$ without any time delay, which further caused variation in the accretion rate.

For Type-B QPOs detected during SIMS (SIMS1, SIMS2, and SIMS3 segment), we found that the variations in coronal heating rate and accretion rate, along with a time delay, can explain the observed temporal features of QPOs. However, as the observed time lags swung between positive and negative, the modeled phase lag also came out to be changing between positive and negative. The associated time delays are greater than those for Type-C QPOs. This could be due to corona contraction or a change in the orientation of the corona from horizontal to vertical \citep{2023Ma}. We also found that this same model can describe the observed rms and lag for an aperiodic hump in SIMS4, suggesting that the same combination of variations can give rise to aperiodic humps also. 

Lastly, for the HS segment, we required variations in $\dot{M}$, $\dot{H}$, and $f_{sc}$ along with some phase delays between them to describe the observed temporal features. Table \ref{tab:Physical_parameter_NICER} gives information about the variations required to describe different variability features. We note here that the nature of lag for this particular source kept swinging between positive and negative between different spectral states as well. It can imply a continuous change in the dynamic origin of Type-B QPOs between the disk and corona. This was also observed by \cite{garg2022energy} for just Type-C QPOs in MAXI J1535-571 during its HIMS and thereby pointed to an underlying resonant phenomenon giving rise to different origins of variability, but there could be some complex scenarios. 

Moreover, the time lags for all segments are in the range of ms, which is significantly smaller than the time scale of viscous or mass inflows. \citet{2020Misra} suggested that there could be a propagation of sound waves over ms scales and can give rise to variability in X-ray binaries. It is thereby possible that the perturbation travels from the disk to the corona over sound scales and gives rise to hard lags whereas the perturbations travel from the corona to the disk over a light travel time scales. However, our conclusions are based on the model fitted with the energy-dependent rms and time lag of this particular source hence it cannot be generalized. The approach applied here has the advantage of getting information on distinctive radiative components responsible for the variation leading to the rise of QPOs. 


\section{ACKNOWLEDGEMENT}
We would like express our gratitude to the High Energy Astrophysics Science Archive Research Centre (HEASARC) for providing the software and data utilised n this project. We are grateful to the LAXPC Payload Operation Center (POC) at IUCAA, Pune, the SXT POC at TIFR, Mumbai, and XRT/NICER for providing the data, the necessary tools, and the analysis guidelines. This research has made use of the MAXI data provided by RIKEN, JAXA and the MAXI team. A.P. and B.S. acknowledge the hospitality and facilities provided by the Inter-University Centre for Astronomy and Astrophysics (IUCAA), Pune, where part of this work was carried out. B.S. would also like to acknowledge the IUCAA visiting associateship program. We thank the anonymous referee for the insightful comments and suggestions.

\section{DATA AVAILABILITY}
The archived data utilized in this project are available at Astrosat-ISSDC website (\url{http://astrobrowse.issdc.gov.in/astro_archive/archive}) and at the NASA’s HEASARC website (\url{https://heasarc.gsfc.nasa.gov/cgi-bin/W3Browse/w3browse.pl}).

\begin{appendix}

\counterwithin{table}{section}
\restartappendixnumbering
\renewcommand\thefigure{\thesection\arabic{figure}} 
\renewcommand\thetable{\thesection\arabic{table}} 
\section{PDS parameters.}

From Figure \ref{fig:PDS_seg_1_7}, it can be seen that the sub-harmonic, fundamental, and harmonic frequencies are not detected in all segments. We have provided details of all the Lorentzian components used to model the power spectra in Table \ref{tab:A1}. Additionally, in Table \ref{tab:A1}, we have highlighted the sub-harmonic frequencies (when detected) in blue at $f_3$ (Hz), and the fundamental frequencies (or broad components in the HS and SIMS4 segment), used to calculate the rms and lag in red at $f_2$ (Hz).

\begin{table}[htbp]
	\hspace{-0.7cm}
	\centering
	\small
	\setlength{\tabcolsep}{2pt} 
	\renewcommand{\arraystretch}{1.7} 
	\caption{Best-fitting Lorentzian parameters of the PDS for all segments.}
	\label{tab:A1}
	\begin{tabular}{cccccccc}
		\hline \hline
		\textbf{Components} & \textbf{HS} & \textbf{HIMS} & \textbf{SIMS1} & \textbf{SIMS2} & \textbf{SIMS3} & \textbf{SIMS4} & \textbf{HSS} \\ \hline \hline
		$ f_1$ \textbf{(Hz)}   & $ 0.0^*$ & $ 0.0^*$ & $ 0.0^*$ & $ 0.0^*$ & $ 0.0^*$ & - & $ 0.0^*$ \\ 
		$ \sigma_{1}$  & $ 3.4^{+1.0}_{-0.7}$ & $ 0.02^{+0.01}_{-0.01}$ & $ 0.05^{+0.02}_{-0.01}$ & $ 1.4^{+0.3}_{-0.2}$ & $ 0.07^{+0.02}_{-0.02}$ & - & $ 0.06^{+0.02}_{-0.03}$ \\ 
		$ N_{L1}$ ($10^{-3}$)   & $ 15.0^{+4.0}_{-4.0}$ & $ 0.5^{+0.1}_{-0.1}$ & $ 0.4^{+0.1}_{-0.1}$ & $ 1.1^{+0.1}_{-0.1}$ & $ 0.5^{+0.1}_{-0.1}$ & -  & $ 0.37^{+0.05}_{-0.09}$ \\ \hline 
		$ f_2$ \textbf{(Hz)} & \red{$ 0.25^{+0.02}_{-0.02}$} & \red{$ 5.74^{+0.02}_{-0.02}$} & \red{$ 5.7^{+0.3}_{-0.2}$} & \red{$ 6.6^{+0.1}_{-0.2}$} & \red{$ 5.17^{+0.05}_{-0.05}$} & \red{$ 0.03^{+0.01}_{-0.03}$} & - \\ 
		$ \sigma_{2}$ & $ 0.20^{+0.09}_{-0.07}$ & $ 1.2^{+0.1}_{-0.1}$ & $ 0.7^{+1.0}_{-0.5}$ & $ 1.3^{+0.3}_{-0.7}$ & $ 0.5^{+0.1}_{-0.1}$ & $ 0.03^{+0.04}_{-0.02}$ & - \\ 
		$ N_{L2}$ ($10^{-3}$)  & $ 20^{+10}_{-10}$  & $ 14.0^{+0.1}_{-0.1}$ & $ 0.04^{+0.03}_{-0.02}$ & $ 0.9^{+0.2}_{-0.7}$ & $ 0.18^{+0.03}_{-0.03}$ & $ 0.09^{+0.18}_{-0.07}$ &-  \\ \hline
		$ f_3$ \textbf{(Hz)} & - & \blue{$ 2.84^{+0.04}_{-0.04}$} & - & \blue{$ 3.26^{+0.06}_{-0.06}$}  & \blue{$ 2.8^{+0.1}_{-0.1}$} & - & -  \\ 
		$\sigma_{3}$ & - & $ 1.2^{+0.2}_{-0.1}$ &- & $ 2.1^{+0.2}_{-0.2}$ & $ 0.8^{+0.3}_{-0.2}$ & -  & - \\ 
		$ N_{L3}$ ($10^{-3}$) & - & $ 4.0^{+1.0}_{-1.0}$ & - & $ 1.4^{+0.1}_{-0.1}$ & $ 0.16^{+0.05}_{-0.04}$ & - & - \\ \hline
		$ f_4$ \textbf{(Hz)} & $ 0.59^{+0.08}_{-0.10}$ & $ 0.31^{+0.07}_{-0.10}$ & $ 0.06^{+0.09}_{-0.06}$ & $ 7.9^{+0.4}_{-1.0}$ & $ 0.0^*$ & $ 0.16^{+0.01}_{-0.01}$ & $ 0.14^{+0.07}_{-0.14}$ \\ 
		$\sigma_{4}$ & $ 0.6^{+0.2}_{-0.2}$ & $ 2.6^{+0.3}_{-0.3}$ & $ 0.41^{+0.12}_{-0.06}$ & $ 1.8^{+1.2}_{-0.9}$ & $ 1.6^{+0.7}_{-0.5}$ & $ 0.02^{+0.04}_{-0.02}$ & $ 0.4^{+0.2}_{-0.1}$  \\ 
		$ N_{L4}$ ($10^{-3}$) & $ 11.0^{+6.0}_{-4.0}$ & $ 13.0^{+1.0}_{-1.0}$ & $ 0.3^{+0.1}_{-0.1}$  & $ 0.3^{+0.7}_{-0.2}$ & $ 0.3^{+0.1}_{-0.1}$ & $ 0.02^{+0.02}_{-0.01}$ & $ 0.2^{+0.1}_{-0.1}$ \\ \hline
		$ f_5$ (Hz) & $ 0.02^{+0.02}_{-0.02}$ & $ 8.8^{+0.4}_{-0.5}$ &$ 0.5^{+1.2}_{-0.5}$& $ 1.6^{+0.1}_{-0.1}$ & - & $ 0.29^{+0.05}_{-0.29}$ & $ 0.65^{+0.03}_{-0.02}$  \\ 
		$\sigma_{5}$ & $ 0.14^{+0.13}_{-0.07}$ & $ 7.8^{+0.7}_{-0.6}$ &  $ 3.7^{+2.4}_{-1.4}$ & $ 1.0^{+0.7}_{-0.4}$ & - & $ 0.2^{+0.4}_{-0.1}$ & $ 0.12^{+0.17}_{-0.08}$ \\ 
		$ N_{L5}$ ($10^{-3}$) & $ 8.0^{+7.0}_{-4.0}$ & $ 6.0^{+1.0}_{-1.0}$ & $ 0.2^{+0.1}_{-0.1}$ & $ 0.2^{+0.2}_{-0.1}$ & - & $ 0.05^{+0.13}_{-0.03}$ & $ 0.03^{+0.02}_{-0.01}$  \\ \hline
		$ f_6$ \textbf{(Hz)} & - & $ 0.50^{+0.03}_{-0.06}$  & - & $ 12.0^{+0.7}_{-0.4}$ & - & - & $ 0.0^*$ \\ 
		$\sigma_{6}$ & - & $ 0.1^{+0.2}_{-0.1}$ & - & $ 0.03^{+16}_{-0.02}$ & - & - & $ 8.1^{+10}_{-4.0}$ \\ 
		$ N_{L6}$ ($10^{-3}$) & -  & $ 0.1^{+0.2}_{-0.1}$ & -& $ 0.28^{+0.03}_{-0.04}$  & - & - & $ 0.15^{+0.05}_{-0.05}$ \\ \hline
		$ PL_{Index}$ & - & - & - & $ 1.6^{+0.1}_{-0.1}$  & - & $ 1.2^{+0.1}_{-0.1}$ & - \\ 
		$ PL_{norm}$ ($10^{-3}$) & - & - & - & $ 0.02^{+0.01}_{-0.01}$ & - & $ 0.06^{+0.01}_{-0.02}$ & - \\ \hline \hline
		$\chi^{2}$/dof &  $ 93.77/76$ & $ 141.85/97$ & $ 80.11/68$ & $ 70.82/40$ & $ 44.69/63$ & $ 64.59/48$ & $ 50/40$ \\ \hline \hline
	\end{tabular}
	\begin{flushleft}
		\textbf{Note:} $^*$ denotes a fixed parameter during the fitting. $f_2$ (Hz) shows the QPO/peak frequency component which is used to calculate rms/lag, $f_3$ (Hz) is the sub harmonic component of the QPO present in HIMS, SIMS2 and SIMS3. $\sigma_{1-6}$ and $N_{L1-6}$ represents the Lorentzian widths, and their normalizations receptively. $PL_{Index}$ and $PL_{norm}$ represents the components of power law used in fitting broad feature in SIMS2 and SIMS4.
	\end{flushleft}
\end{table}

\end{appendix}

\bibliography{sample631}{}

\begin{thebibliography}{}
\expandafter\ifx\csname natexlab\endcsname\relax\def\natexlab#1{#1}\fi
\providecommand{\url}[1]{\href{#1}{#1}}
\providecommand{\dodoi}[1]{doi:~\href{http://doi.org/#1}{\nolinkurl{#1}}}
\providecommand{\doeprint}[1]{\href{http://ascl.net/#1}{\nolinkurl{http://ascl.net/#1}}}
\providecommand{\doarXiv}[1]{\href{https://arxiv.org/abs/#1}{\nolinkurl{https://arxiv.org/abs/#1}}}

\bibitem[{{Adegoke} {et~al.}(2024){Adegoke}, {Garc{\'\i}a}, {Connors}, {Ding},
  {Mastroserio}, {Steiner}, {Ingram}, {Harrison}, {Tomsick}, {Kara},
  {Mehdipour}, {Fukumura}, {Stern}, {Ubach}, \& {Lucchini}}]{2024Adegoke}
{Adegoke}, O.~K., {Garc{\'\i}a}, J.~A., {Connors}, R. M.~T., {et~al.} 2024,
  \apj, 977, 26, \dodoi{10.3847/1538-4357/ad82e9}

\bibitem[{{Agrawal}(2006)}]{2006AdSpR..38.2989A}
{Agrawal}, P.~C. 2006, Advances in Space Research, 38, 2989,
  \dodoi{10.1016/j.asr.2006.03.038}

\bibitem[{{Altamirano} \& {Belloni}(2012)}]{2012ApJ...747L...4A}
{Altamirano}, D., \& {Belloni}, T. 2012, \apjl, 747, L4,
  \dodoi{10.1088/2041-8205/747/1/L4}

\bibitem[{{Antia} {et~al.}(2017){Antia}, {Yadav}, {Agrawal}, {Verdhan Chauhan},
  {Manchanda}, {Chitnis}, {Paul}, {Dedhia}, {Shah}, {Gujar}, {Katoch},
  {Kurhade}, {Madhwani}, {Manojkumar}, {Nikam}, {Pandya}, {Parmar}, {Pawar},
  {Pahari}, {Misra}, {Navalgund}, {Pandiyan}, {Sharma}, \&
  {Subbarao}}]{2017ApJS..231...10A}
{Antia}, H.~M., {Yadav}, J.~S., {Agrawal}, P.~C., {et~al.} 2017, \apjs, 231,
  10, \dodoi{10.3847/1538-4365/aa7a0e}

\bibitem[{Arzoumanian {et~al.}(2014)Arzoumanian, Gendreau, Baker, Cazeau,
  Hestnes, Kellogg, Kenyon, Kozon, Liu, Manthripragada,
  {et~al.}}]{arzoumanian2014neutron}
Arzoumanian, Z., Gendreau, K., Baker, C., {et~al.} 2014, in Space Telescopes
  and Instrumentation 2014: Ultraviolet to Gamma Ray, Vol. 9144, SPIE, 579--587

\bibitem[{{Bellavita} {et~al.}(2022){Bellavita}, {Garc{\'\i}a}, {M{\'e}ndez},
  \& {Karpouzas}}]{bellavita2022vkompth}
{Bellavita}, C., {Garc{\'\i}a}, F., {M{\'e}ndez}, M., \& {Karpouzas}, K. 2022,
  \mnras, 515, 2099, \dodoi{10.1093/mnras/stac1922}

\bibitem[{{Belloni} {et~al.}(2005{\natexlab{a}}){Belloni}, {Homan}, {Casella},
  {van der Klis}, {Nespoli}, {Lewin}, {Miller}, \&
  {M{\'e}ndez}}]{2005A&A...440..207B}
{Belloni}, T., {Homan}, J., {Casella}, P., {et~al.} 2005{\natexlab{a}}, \aap,
  440, 207, \dodoi{10.1051/0004-6361:20042457}

\bibitem[{{Belloni} {et~al.}(2005{\natexlab{b}}){Belloni}, {Homan}, {Casella},
  {van der Klis}, {Nespoli}, {Lewin}, {Miller}, \& {M{\'e}ndez}}]{2005Belloni}
---. 2005{\natexlab{b}}, \aap, 440, 207, \dodoi{10.1051/0004-6361:20042457}

\bibitem[{{Belloni} \& {Altamirano}(2013)}]{2013MNRAS.432...10B}
{Belloni}, T.~M., \& {Altamirano}, D. 2013, \mnras, 432, 10,
  \dodoi{10.1093/mnras/stt500}

\bibitem[{{Belloni} \& {Motta}(2016)}]{2016ASSL..440...61B}
{Belloni}, T.~M., \& {Motta}, S.~E. 2016, in Astrophysics and Space Science
  Library, Vol. 440, Astrophysics of Black Holes: From Fundamental Aspects to
  Latest Developments, ed. C.~{Bambi}, 61, \dodoi{10.1007/978-3-662-52859-4_2}

\bibitem[{{Belloni} {et~al.}(2011){Belloni}, {Motta}, \&
  {Mu{\~n}oz-Darias}}]{2011BASI...39..409B}
{Belloni}, T.~M., {Motta}, S.~E., \& {Mu{\~n}oz-Darias}, T. 2011, Bulletin of
  the Astronomical Society of India, 39, 409, \dodoi{10.48550/arXiv.1109.3388}

\bibitem[{{Bhattacharyya} {et~al.}(2021){Bhattacharyya}, {Singh}, {Stewart},
  {Chandra}, {Dewangan}, {Kamble}, {Vishwakarma}, {Koyande}, \&
  {Chitnis}}]{bhattacharyya2021science}
{Bhattacharyya}, S., {Singh}, K.~P., {Stewart}, G., {et~al.} 2021, Journal of
  Astrophysics and Astronomy, 42, 17, \dodoi{10.1007/s12036-020-09678-z}

\bibitem[{{Bhattacherjee} {et~al.}(2024{\natexlab{a}}){Bhattacherjee}, {Nath},
  {Sarkar}, {Beri}, {Chattopadhyay}, {Bhulla}, \& {Misra}}]{2024Sree}
{Bhattacherjee}, S., {Nath}, A., {Sarkar}, B., {et~al.} 2024{\natexlab{a}},
  \apj, 971, 154, \dodoi{10.3847/1538-4357/ad583d}

\bibitem[{{Bhattacherjee} {et~al.}(2024{\natexlab{b}}){Bhattacherjee},
  {Pradhan}, \& {Sarkar}}]{2024GX9+9}
{Bhattacherjee}, S., {Pradhan}, A., \& {Sarkar}, B. 2024{\natexlab{b}}, Physics
  Frontiers, Vol-I, 18, \dodoi{10.48550/arXiv.2409.11721}

\bibitem[{{Buckley} {et~al.}(2021){Buckley}, {Brink}, {Charles}, \&
  {Groenewald}}]{2021ATel14597....1B}
{Buckley}, D.~A.~H., {Brink}, J., {Charles}, P.~A., \& {Groenewald}, D. 2021,
  The Astronomer's Telegram, 14597, 1

\bibitem[{{Buisson} {et~al.}(2019){Buisson}, {Fabian}, {Barret}, {F{\"u}rst},
  {Gandhi}, {Garc{\'\i}a}, {Kara}, {Madsen}, {Miller}, {Parker}, {Shaw},
  {Tomsick}, \& {Walton}}]{2019MNRAS.490.1350B}
{Buisson}, D.~J.~K., {Fabian}, A.~C., {Barret}, D., {et~al.} 2019, \mnras, 490,
  1350, \dodoi{10.1093/mnras/stz2681}

\bibitem[{Casella {et~al.}(2004)Casella, Belloni, Homan, \&
  Stella}]{casella2004study}
Casella, P., Belloni, T., Homan, J., \& Stella, L. 2004, Astronomy \&
  Astrophysics, 426, 587

\bibitem[{{Chakrabarti} \& {Manickam}(2000)}]{2000ApJCha}
{Chakrabarti}, S.~K., \& {Manickam}, S.~G. 2000, \apjl, 531, L41,
  \dodoi{10.1086/312512}

\bibitem[{{Chand} {et~al.}(2022){Chand}, {Dewangan}, {Thakur}, {Tripathi}, \&
  {Agrawal}}]{2022ApJ...933...69C}
{Chand}, S., {Dewangan}, G.~C., {Thakur}, P., {Tripathi}, P., \& {Agrawal},
  V.~K. 2022, \apj, 933, 69, \dodoi{10.3847/1538-4357/ac7154}

\bibitem[{{Coughenour} {et~al.}(2023){Coughenour}, {Tomsick}, {Mastroserio},
  {Steiner}, {Connors}, {Jiang}, {Hare}, {Shaw}, {Ludlam}, {Fabian},
  {Garc{\'\i}a}, \& {Coley}}]{2023ApJ...949...70C}
{Coughenour}, B.~M., {Tomsick}, J.~A., {Mastroserio}, G., {et~al.} 2023, \apj,
  949, 70, \dodoi{10.3847/1538-4357/acc65c}

\bibitem[{{Dhaka} {et~al.}(2024){Dhaka}, {Misra}, {Jain}, \&
  {Yadav}}]{2024Dhaka}
{Dhaka}, R., {Misra}, R., {Jain}, P., \& {Yadav}, J.~S. 2024, \apj, 974, 90,
  \dodoi{10.3847/1538-4357/ad67e4}

\bibitem[{{Dhaka} {et~al.}(2023){Dhaka}, {Misra}, {Yadav}, \&
  {Jain}}]{2023MNRAS.524.2721D}
{Dhaka}, R., {Misra}, R., {Yadav}, J.~S., \& {Jain}, P. 2023, \mnras, 524,
  2721, \dodoi{10.1093/mnras/stad2075}

\bibitem[{{Dunn} {et~al.}(2010){Dunn}, {Fender}, {K{\"o}rding}, {Belloni}, \&
  {Cabanac}}]{2010MNRAS.403...61D}
{Dunn}, R.~J.~H., {Fender}, R.~P., {K{\"o}rding}, E.~G., {Belloni}, T., \&
  {Cabanac}, C. 2010, \mnras, 403, 61, \dodoi{10.1111/j.1365-2966.2010.16114.x}

\bibitem[{{Espinasse} {et~al.}(2021){Espinasse}, {Carotenuto}, {Tremou},
  {Corbel}, {Fender}, {Woudt}, \& {Miller-Jones}}]{2021ATel14607....1E}
{Espinasse}, M., {Carotenuto}, F., {Tremou}, E., {et~al.} 2021, The
  Astronomer's Telegram, 14607, 1

\bibitem[{{Feng} {et~al.}(2022){Feng}, {Zhao}, {Li}, {Gou}, {Jia}, {Liao}, \&
  {Wang}}]{2022MNRAS.516.2074F}
{Feng}, Y., {Zhao}, X., {Li}, Y., {et~al.} 2022, \mnras, 516, 2074,
  \dodoi{10.1093/mnras/stac1868}

\bibitem[{{Gandhi} {et~al.}(2011){Gandhi}, {Blain}, {Russell}, {Casella},
  {Malzac}, {Corbel}, {D'Avanzo}, {Lewis}, {Markoff}, {Cadolle Bel}, {Goldoni},
  {Wachter}, {Khangulyan}, \& {Mainzer}}]{2011ApJ...740L..13G}
{Gandhi}, P., {Blain}, A.~W., {Russell}, D.~M., {et~al.} 2011, \apjl, 740, L13,
  \dodoi{10.1088/2041-8205/740/1/L13}

\bibitem[{{Garc{\'\i}a} {et~al.}(2022){Garc{\'\i}a}, {Karpouzas}, {M{\'e}ndez},
  {Zhang}, {Zhang}, {Belloni}, \& {Altamirano}}]{garcia2022evolving}
{Garc{\'\i}a}, F., {Karpouzas}, K., {M{\'e}ndez}, M., {et~al.} 2022, \mnras,
  513, 4196, \dodoi{10.1093/mnras/stac1202}

\bibitem[{{Garg} {et~al.}(2020){Garg}, {Misra}, \& {Sen}}]{garg2020identifying}
{Garg}, A., {Misra}, R., \& {Sen}, S. 2020, \mnras, 498, 2757,
  \dodoi{10.1093/mnras/staa2506}

\bibitem[{{Garg} {et~al.}(2022){Garg}, {Misra}, \& {Sen}}]{garg2022energy}
---. 2022, \mnras, 514, 3285, \dodoi{10.1093/mnras/stac1490}

\bibitem[{{Gendreau} \& {Arzoumanian}(2017)}]{gendreau2017searching}
{Gendreau}, K., \& {Arzoumanian}, Z. 2017, Nature Astronomy, 1, 895,
  \dodoi{10.1038/s41550-017-0301-3}

\bibitem[{Gendreau {et~al.}(2016)Gendreau, Arzoumanian, Adkins, Albert, Anders,
  Aylward, Baker, Balsamo, Bamford, Benegalrao, {et~al.}}]{gendreau2016neutron}
Gendreau, K.~C., Arzoumanian, Z., Adkins, P.~W., {et~al.} 2016, in Space
  telescopes and instrumentation 2016: Ultraviolet to gamma ray, Vol. 9905,
  SPIE, 420--435

\bibitem[{{Gropp} {et~al.}(2021){Gropp}, {Kennea}, {Lien}, {Marshall}, {Page},
  {Palmer}, {Sbarufatti}, {Siegel}, \& {Tohuvavohu}}]{2021ATel14591....1G}
{Gropp}, J.~D., {Kennea}, J.~A., {Lien}, A.~Y., {et~al.} 2021, The Astronomer's
  Telegram, 14591, 1

\bibitem[{{Homan} \& {Belloni}(2005)}]{2005Ap&SS.300..107H}
{Homan}, J., \& {Belloni}, T. 2005, \apss, 300, 107,
  \dodoi{10.1007/s10509-005-1197-4}

\bibitem[{Homan {et~al.}(2001)Homan, Wijnands, van~der Klis, Belloni, van
  Paradijs, Klein-Wolt, Fender, \& Mendez}]{homan2001correlated}
Homan, J., Wijnands, R., van~der Klis, M., {et~al.} 2001, The Astrophysical
  Journal Supplement Series, 132, 377

\bibitem[{{Homan} {et~al.}(2021){Homan}, {Gendreau}, {Sanna}, {Jaisawal},
  {Buisson}, {Bult}, {Altamirano}, {Neilsen}, \& {Kara}}]{2021ATel14606....1H}
{Homan}, J., {Gendreau}, K.~C., {Sanna}, A., {et~al.} 2021, The Astronomer's
  Telegram, 14606, 1

\bibitem[{{Hosokawa} {et~al.}(2021){Hosokawa}, {Murata}, {Niwano}, {Ito},
  {Takamatsu}, {Imai}, {Sato}, {Takaku}, {Noto}, {Yamaguchi}, {Yatsu}, \&
  {Kawai}}]{2021ATel14594....1H}
{Hosokawa}, R., {Murata}, K.~L., {Niwano}, M., {et~al.} 2021, The Astronomer's
  Telegram, 14594, 1

\bibitem[{{Husain} {et~al.}(2023){Husain}, {Garg}, {Misra}, \&
  {Sen}}]{husain2023investigating}
{Husain}, N., {Garg}, A., {Misra}, R., \& {Sen}, S. 2023, \mnras, 525, 4515,
  \dodoi{10.1093/mnras/stad2481}

\bibitem[{Ingram {et~al.}(2009)Ingram, Done, \& Fragile}]{ingram2009low}
Ingram, A., Done, C., \& Fragile, P.~C. 2009, Monthly Notices of the Royal
  Astronomical Society: Letters, 397, L101

\bibitem[{Jana {et~al.}(2022)Jana, Naik, Jaisawal, Chhotaray, Kumari, \&
  Gupta}]{Jana_2022}
Jana, A., Naik, S., Jaisawal, G.~K., {et~al.} 2022, Monthly Notices of the
  Royal Astronomical Society, 511, 3922–3936, \dodoi{10.1093/mnras/stac315}

\bibitem[{{Kara} {et~al.}(2019){Kara}, {Steiner}, {Fabian}, {Cackett},
  {Uttley}, {Remillard}, {Gendreau}, {Arzoumanian}, {Altamirano}, {Eikenberry},
  {Enoto}, {Homan}, {Neilsen}, \& {Stevens}}]{2019Natur.565..198K}
{Kara}, E., {Steiner}, J.~F., {Fabian}, A.~C., {et~al.} 2019, \nat, 565, 198,
  \dodoi{10.1038/s41586-018-0803-x}

\bibitem[{{Karpouzas} {et~al.}(2021){Karpouzas}, {M{\'e}ndez}, {Garc{\'\i}a},
  {Zhang}, {Altamirano}, {Belloni}, \& {Zhang}}]{karpouzas2021variable}
{Karpouzas}, K., {M{\'e}ndez}, M., {Garc{\'\i}a}, F., {et~al.} 2021, \mnras,
  503, 5522, \dodoi{10.1093/mnras/stab827}

\bibitem[{{Karpouzas} {et~al.}(2020){Karpouzas}, {M{\'e}ndez}, {Ribeiro},
  {Altamirano}, {Blaes}, \& {Garc{\'\i}a}}]{karpouzas2020comptonizing}
{Karpouzas}, K., {M{\'e}ndez}, M., {Ribeiro}, E.~M., {et~al.} 2020, \mnras,
  492, 1399, \dodoi{10.1093/mnras/stz3502}

\bibitem[{{Kumar} \& {Misra}(2014)}]{kumar2014energy}
{Kumar}, N., \& {Misra}, R. 2014, \mnras, 445, 2818,
  \dodoi{10.1093/mnras/stu1946}

\bibitem[{{Kylafis} \& {Belloni}(2015)}]{2015A&A...574A.133K}
{Kylafis}, N.~D., \& {Belloni}, T.~M. 2015, \aap, 574, A133,
  \dodoi{10.1051/0004-6361/201425106}

\bibitem[{{Laor}(1991)}]{1991ApJ...376...90L}
{Laor}, A. 1991, \apj, 376, 90, \dodoi{10.1086/170257}

\bibitem[{{Lee} {et~al.}(2001){Lee}, {Misra}, \& {Taam}}]{lee2001compton}
{Lee}, H.~C., {Misra}, R., \& {Taam}, R.~E. 2001, \apjl, 549, L229,
  \dodoi{10.1086/319171}

\bibitem[{{Lewin} {et~al.}(1997){Lewin}, {van Paradijs}, \& {van den
  Heuvel}}]{1997xrb..book.....L}
{Lewin}, W. H.~G., {van Paradijs}, J., \& {van den Heuvel}, E. P.~J. 1997,
  {X-ray Binaries}

\bibitem[{{Liu} {et~al.}(2007){Liu}, {Taam}, {Meyer-Hofmeister}, \&
  {Meyer}}]{2007Liu}
{Liu}, B.~F., {Taam}, R.~E., {Meyer-Hofmeister}, E., \& {Meyer}, F. 2007, \apj,
  671, 695, \dodoi{10.1086/522619}

\bibitem[{{Ma} {et~al.}(2023){Ma}, {M{\'e}ndez}, {Garc{\'\i}a}, {Sai}, {Zhang},
  \& {Zhang}}]{2023Ma}
{Ma}, R., {M{\'e}ndez}, M., {Garc{\'\i}a}, F., {et~al.} 2023, \mnras, 525, 854,
  \dodoi{10.1093/mnras/stad2284}

\bibitem[{{Maqbool} {et~al.}(2019){Maqbool}, {Mudambi}, {Misra}, {Yadav},
  {Gudennavar}, {Bubbly}, {Rao}, {Jogadand}, {Patil}, {Bhattacharyya}, \&
  {Singh}}]{maqbool2019stochastic}
{Maqbool}, B., {Mudambi}, S.~P., {Misra}, R., {et~al.} 2019, \mnras, 486, 2964,
  \dodoi{10.1093/mnras/stz930}

\bibitem[{{Mata S{\'a}nchez} {et~al.}(2022){Mata S{\'a}nchez},
  {Mu{\~n}oz-Darias}, {C{\'u}neo}, {Armas Padilla}, {S{\'a}nchez-Sierras},
  {Panizo-Espinar}, {Casares}, {Corral-Santana}, \& {Torres}}]{Mata_2022}
{Mata S{\'a}nchez}, D., {Mu{\~n}oz-Darias}, T., {C{\'u}neo}, V.~A., {et~al.}
  2022, \apjl, 926, L10, \dodoi{10.3847/2041-8213/ac502f}

\bibitem[{{Miller}(2007)}]{2007ARA&A..45..441M}
{Miller}, J.~M. 2007, \araa, 45, 441,
  \dodoi{10.1146/annurev.astro.45.051806.110555}

\bibitem[{Mir {et~al.}(2016)Mir, Misra, Pahari, Iqbal, \& Ahmad}]{mir2016model}
Mir, M.~H., Misra, R., Pahari, M., Iqbal, N., \& Ahmad, N. 2016, Monthly
  Notices of the Royal Astronomical Society, 457, 2999

\bibitem[{Misra \& Mandal(2013)}]{misra2013alternating}
Misra, R., \& Mandal, S. 2013, The Astrophysical Journal, 779, 71

\bibitem[{{Misra} {et~al.}(2020){Misra}, {Rawat}, {Yadav}, \&
  {Jain}}]{2020Misra}
{Misra}, R., {Rawat}, D., {Yadav}, J.~S., \& {Jain}, P. 2020, \apjl, 889, L36,
  \dodoi{10.3847/2041-8213/ab6ddc}

\bibitem[{{Misra} {et~al.}(2017){Misra}, {Yadav}, {Verdhan Chauhan}, {Agrawal},
  {Antia}, {Pahari}, {Chitnis}, {Dedhia}, {Katoch}, {Madhwani}, {Manchanda},
  {Paul}, \& {Shah}}]{2017ApJ...Ranjeev}
{Misra}, R., {Yadav}, J.~S., {Verdhan Chauhan}, J., {et~al.} 2017, \apj, 835,
  195, \dodoi{10.3847/1538-4357/835/2/195}

\bibitem[{{Mitsuda} {et~al.}(1984){Mitsuda}, {Inoue}, {Koyama}, {Makishima},
  {Matsuoka}, {Ogawara}, {Shibazaki}, {Suzuki}, {Tanaka}, \&
  {Hirano}}]{1984PASJ...36..741M}
{Mitsuda}, K., {Inoue}, H., {Koyama}, K., {et~al.} 1984, \pasj, 36, 741

\bibitem[{Mitsuda {et~al.}(1984)Mitsuda, Inoue, Koyama, Makishima, Matsuoka,
  Ogawara, Shibazaki, Suzuki, Tanaka, \& Hirano}]{mitsuda1984energy}
Mitsuda, K., Inoue, H., Koyama, K., {et~al.} 1984, Astronomical Society of
  Japan, Publications (ISSN 0004-6264), vol. 36, no. 4, 1984, p. 741-759., 36,
  741

\bibitem[{{Miyamoto} {et~al.}(1995){Miyamoto}, {Kitamoto}, {Hayashida}, \&
  {Egoshi}}]{1995ApJ...442L..13M}
{Miyamoto}, S., {Kitamoto}, S., {Hayashida}, K., \& {Egoshi}, W. 1995, \apjl,
  442, L13, \dodoi{10.1086/187804}

\bibitem[{{Molteni} {et~al.}(1999){Molteni}, {T{\'o}th}, \&
  {Kuznetsov}}]{1999ApJMol}
{Molteni}, D., {T{\'o}th}, G., \& {Kuznetsov}, O.~A. 1999, \apj, 516, 411,
  \dodoi{10.1086/307079}

\bibitem[{{Motta}(2016)}]{2016AN....337..398M}
{Motta}, S.~E. 2016, Astronomische Nachrichten, 337, 398,
  \dodoi{10.1002/asna.201612320}

\bibitem[{{Nandi} {et~al.}(2012){Nandi}, {Debnath}, {Mandal}, \&
  {Chakrabarti}}]{2012A&A...542A..56N}
{Nandi}, A., {Debnath}, D., {Mandal}, S., \& {Chakrabarti}, S.~K. 2012, \aap,
  542, A56, \dodoi{10.1051/0004-6361/201117844}

\bibitem[{{Narayan} \& {Yi}(1995)}]{1995Narayan}
{Narayan}, R., \& {Yi}, I. 1995, \apj, 452, 710, \dodoi{10.1086/176343}

\bibitem[{{Parker} {et~al.}(2015){Parker}, {Tomsick}, {Miller}, {Yamaoka},
  {Lohfink}, {Nowak}, {Fabian}, {Alston}, {Boggs}, {Christensen}, {Craig},
  {F{\"u}rst}, {Gandhi}, {Grefenstette}, {Grinberg}, {Hailey}, {Harrison},
  {Kara}, {King}, {Stern}, {Walton}, {Wilms}, \& {Zhang}}]{2015ApJParker}
{Parker}, M.~L., {Tomsick}, J.~A., {Miller}, J.~M., {et~al.} 2015, \apj, 808,
  9, \dodoi{10.1088/0004-637X/808/1/9}

\bibitem[{{Rout} {et~al.}(2021){Rout}, {M{\'e}ndez}, {Belloni}, \&
  {Vadawale}}]{2021MNRAS.505.1213R}
{Rout}, S.~K., {M{\'e}ndez}, M., {Belloni}, T.~M., \& {Vadawale}, S. 2021,
  \mnras, 505, 1213, \dodoi{10.1093/mnras/stab1341}

\bibitem[{{Serino} {et~al.}(2021){Serino}, {Negoro}, {Nakajima}, {Kobayashi},
  {Asakura}, {Seino}, {Mihara}, {Tamagawa}, {Matsuoka}, {Sakamoto}, {Sugita},
  {Komachi}, {Yoshida}, {Tsuboi}, {Iwakiri}, {Kawai}, {Okamoto}, {Kitakoga},
  {Shidatsu}, {Kawai}, {Adachi}, {Niwano}, {Hosokawa}, {Nakahira}, {Sugawara},
  {Ueno}, {Tomida}, {Ishikawa}, {Tominaga}, {Nagatsuka}, {Ueda}, {Yamada},
  {Ogawa}, {Setoguchi}, {Yoshitake}, {Goto}, {Uematsu}, {Tsunemi}, {Yamauchi},
  {Nonaka}, {Sato}, {Hatsuda}, {Fukuoka}, {Kawamuro}, {Yamaoka}, {Kawakubo}, \&
  {Sugizaki}}]{2021ATel14587....1S}
{Serino}, M., {Negoro}, H., {Nakajima}, M., {et~al.} 2021, The Astronomer's
  Telegram, 14587, 1

\bibitem[{{Shidatsu} {et~al.}(2022){Shidatsu}, {Kobayashi}, {Negoro},
  {Iwakiri}, {Nakahira}, {Ueda}, {Mihara}, {Enoto}, {Gendreau}, {Arzoumanian},
  {Pope}, {Trout}, {Okajima}, \& {Soong}}]{Shidatsu2022}
{Shidatsu}, M., {Kobayashi}, K., {Negoro}, H., {et~al.} 2022, \apj, 927, 151,
  \dodoi{10.3847/1538-4357/ac517b}

\bibitem[{{Singh}(2022)}]{2022hxga.book...83S}
{Singh}, K.~P. 2022, in Handbook of X-ray and Gamma-ray Astrophysics, 83,
  \dodoi{10.1007/978-981-16-4544-0_31-1}

\bibitem[{{Singh} {et~al.}(2017){Singh}, {Stewart}, {Westergaard},
  {Bhattacharayya}, {Chandra}, {Chitnis}, {Dewangan}, {Kothare}, {Mirza},
  {Mukerjee}, {Navalkar}, {Shah}, {Abbey}, {Beardmore}, {Kotak}, {Kamble},
  {Vishwakarama}, {Pathare}, {Risbud}, {Koyande}, {Stevenson}, {Bicknell},
  {Crawford}, {Hansford}, {Peters}, {Sykes}, {Agarwal}, {Sebastian},
  {Rajarajan}, {Nagesh}, {Narendra}, {Ramesh}, {Rai}, {Navalgund}, {Sarma},
  {Pandiyan}, {Subbarao}, {Gupta}, {Thakkar}, {Singh}, \&
  {Bajpai}}]{2017JApA...38...29S}
{Singh}, K.~P., {Stewart}, G.~C., {Westergaard}, N.~J., {et~al.} 2017, Journal
  of Astrophysics and Astronomy, 38, 29, \dodoi{10.1007/s12036-017-9448-7}

\bibitem[{{Singh} {et~al.}(2021){Singh}, {Girish}, {Tiwari}, {Barrett},
  {Buckley}, {Potter}, {Schlegel}, {Rana}, \&
  {Stewart}}]{singh2021observations}
{Singh}, K.~P., {Girish}, V., {Tiwari}, J., {et~al.} 2021, Journal of
  Astrophysics and Astronomy, 42, 83, \dodoi{10.1007/s12036-021-09756-w}

\bibitem[{{Steiner} {et~al.}(2021){Steiner}, {Ubach}, {Tomsick}, {Coughenour},
  \& {Homan}}]{2021ATel14994....1S}
{Steiner}, J.~F., {Ubach}, S., {Tomsick}, J.~A., {Coughenour}, B., \& {Homan},
  J. 2021, The Astronomer's Telegram, 14994, 1

\bibitem[{Stella {et~al.}(1999)Stella, Vietri, \&
  Morsink}]{stella1999correlations}
Stella, L., Vietri, M., \& Morsink, S.~M. 1999, The Astrophysical Journal, 524,
  L63

\bibitem[{{Sunyaev} \& {Titarchuk}(1980)}]{1980A&A....86..121S}
{Sunyaev}, R.~A., \& {Titarchuk}, L.~G. 1980, \aap, 86, 121

\bibitem[{{Tanenia} {et~al.}(2024){Tanenia}, {Garg}, {Misra}, \&
  {Sen}}]{2024Hitesh}
{Tanenia}, H., {Garg}, A., {Misra}, R., \& {Sen}, S. 2024, \apj, 975, 190,
  \dodoi{10.3847/1538-4357/ad7d8b}

\bibitem[{{Titarchuk}(1994)}]{1994ApJ...434..570T}
{Titarchuk}, L. 1994, \apj, 434, 570, \dodoi{10.1086/174760}

\bibitem[{Titarchuk \& Fiorito(2004)}]{titarchuk2004spectral}
Titarchuk, L., \& Fiorito, R. 2004, The Astrophysical Journal, 612, 988

\bibitem[{{Titarchuk} \& {Osherovich}(1999)}]{titarchuk1999correlations}
{Titarchuk}, L., \& {Osherovich}, V. 1999, \apjl, 518, L95,
  \dodoi{10.1086/312083}

\bibitem[{{van der Klis}(2005)}]{2005AN....326..798V}
{van der Klis}, M. 2005, Astronomische Nachrichten, 326, 798,
  \dodoi{10.1002/asna.200510416}

\bibitem[{Wijnands {et~al.}(1999)Wijnands, Homan, \& van~der
  Klis}]{wijnands1999complex}
Wijnands, R., Homan, J., \& van~der Klis, M. 1999, The Astrophysical Journal,
  526, L33

\bibitem[{{Wood} {et~al.}(2023){Wood}, {Miller-Jones}, {Bahramian}, {Tingay},
  {Russell}, {Tetarenko}, {Altamirano}, {Belloni}, {Carotenuto}, {Ceccobello},
  {Corbel}, {Espinasse}, {Fender}, {K{\"o}rding}, {Migliari}, {Russell},
  {Sarazin}, {Sivakoff}, {Soria}, \& {Tudose}}]{2023MNRAS.522...70W}
{Wood}, C.~M., {Miller-Jones}, J.~C.~A., {Bahramian}, A., {et~al.} 2023,
  \mnras, 522, 70, \dodoi{10.1093/mnras/stad939}

\bibitem[{{Xu} \& {Harrison}(2021)}]{2021ATel14609....1X}
{Xu}, Y., \& {Harrison}, F. 2021, The Astronomer's Telegram, 14609, 1

\bibitem[{{Yadav} {et~al.}(2016{\natexlab{a}}){Yadav}, {Agrawal}, {Antia},
  {Chauhan}, {Dedhia}, {Katoch}, {Madhwani}, {Manchanda}, {Misra}, {Pahari},
  {Paul}, \& {Shah}}]{2016SPIE.9905E..1DY}
{Yadav}, J.~S., {Agrawal}, P.~C., {Antia}, H.~M., {et~al.} 2016{\natexlab{a}},
  in Society of Photo-Optical Instrumentation Engineers (SPIE) Conference
  Series, Vol. 9905, Space Telescopes and Instrumentation 2016: Ultraviolet to
  Gamma Ray, ed. J.-W.~A. {den Herder}, T.~{Takahashi}, \& M.~{Bautz}, 99051D,
  \dodoi{10.1117/12.2231857}

\bibitem[{{Yadav} {et~al.}(2016{\natexlab{b}}){Yadav}, {Misra}, {Verdhan
  Chauhan}, {Agrawal}, {Antia}, {Pahari}, {Dedhia}, {Katoch}, {Madhwani},
  {Manchanda}, {Paul}, {Shah}, \& {Ishwara-Chandra}}]{2016ApJ...833...27Y}
{Yadav}, J.~S., {Misra}, R., {Verdhan Chauhan}, J., {et~al.}
  2016{\natexlab{b}}, \apj, 833, 27, \dodoi{10.3847/0004-637X/833/1/27}

\bibitem[{{Zdziarski} {et~al.}(2020){Zdziarski}, {Szanecki}, {Poutanen},
  {Gierli{\'n}ski}, \& {Biernacki}}]{zdziarski2020spectral}
{Zdziarski}, A.~A., {Szanecki}, M., {Poutanen}, J., {Gierli{\'n}ski}, M., \&
  {Biernacki}, P. 2020, \mnras, 492, 5234, \dodoi{10.1093/mnras/staa159}

\bibitem[{{Zhang} {et~al.}(2024){Zhang}, {Bambi}, {Liu}, {Jiang}, {Shi},
  {Zhang}, {Young}, {Tomsick}, {Coughenour}, \& {Zhou}}]{zhang2024variable}
{Zhang}, Z., {Bambi}, C., {Liu}, H., {et~al.} 2024, \apj, 975, 22,
  \dodoi{10.3847/1538-4357/ad7b29}

\bibitem[{{Zhu} {et~al.}(2023){Zhu}, {Chen}, \& {Wang}}]{2023MNRAS.523.4394Z}
{Zhu}, H., {Chen}, X., \& {Wang}, W. 2023, \mnras, 523, 4394,
  \dodoi{10.1093/mnras/stad1656}

\end{thebibliography}
\bibliographystyle{aasjournal}

\end{document}